\documentclass[]{aastex63}
\pdfoutput=1

\newcommand{\acen}{$\alpha$ Cen}
\newcommand{\mearth}{M$_\oplus$}

\accepted{\today}

\shorttitle{alpha Cen Astrometry}
\shortauthors{Akeson et al.}

\graphicspath{{./}{figures/}}

\begin{document}

\title{Precision Millimeter Astrometry of the $\alpha$ Centauri AB System
}

\correspondingauthor{Rachel Akeson}

\author[0000-0001-9674-1564]{Rachel Akeson}
\affiliation{NASA Exoplanet Science Institute \\
Caltech/IPAC, \\
Pasadena, CA, 91125, USA}
\email{rla@ipac.caltech.edu}

\author[0000-0002-5627-5471]{Charles Beichman}
\affiliation{NASA Exoplanet Science Institute, IPAC\\
Jet Propulsion Laboratory, \\
California Institute of Technology, \\
Pasadena, CA, 91125, USA}
\email{chas@ipac.caltech.edu}

\author[0000-0003-0626-1749]{Pierre Kervella}
\affiliation{LESIA, Observatoire de Paris, Universit\'e PSL, CNRS, \\
Sorbonne Universit\'e, Universit\'e de Paris, \\
5 Place Jules Janssen, 92195 Meudon, France}
\email{pierre.kervella@observatoiredeparis.psl.eu}

\author[0000-0002-9036-2747]{Edward Fomalont}
\affiliation{National Radio Astronomy Observatory, \\
Charlottesville, VA, 22903, USA} 
\affiliation{ALMA, Vitacura, Santiago, Chile}
\email{efomalon@nrao.edu}

\author[0000-0003-2852-3279]{G. Fritz Benedict}
\affiliation{McDonald Observatory, \\
University of Texas, \\
Austin, TX 78712}
\email{fritz@astro.as.utexas.edu}

\begin{abstract}

$\alpha$ Centauri A is the closest solar-type star to the Sun and offers the best opportunity to find and ultimately to characterize an Earth-sized planet located in its Habitable Zone (HZ). Here we describe initial results from an ALMA program to search for planets in the \acen\ AB system using differential astrometry at millimeter wavelengths. Our initial results include new absolute astrometric measurements of the proper motion, orbital motion, and parallax of the \acen\ system. These lead to an improved knowledge of the physical properties of both \acen\ A and B. Our estimates of ALMA's relative astrometric precision suggest that we will ultimately be sensitive to planets of a few 10s of Earth mass in orbits from 1-3 AU, where stable orbits are thought to exist. 

\end{abstract}

\section{Introduction}

At a distance of 1.34 pc, \acen\ A is an ideal target for exoplanet searches and is 2.7 times closer than the next most favorable G star, $\tau$ Ceti. \acen\ A's luminosity of 1.5 L$_\odot$ \citep{Thevenin2002} puts the center of its HZ at a physical separation of 1.2 AU which corresponds to an angular separation of 0\farcs9. \acen\ A is thus an attractive target for direct imaging searches for planets using ground or space based observatories \citep{Kasper2019, Beichman2020, Wagner2020}. Current precision radial velocity (PRV) observations \citep{Zhao2018} constrain the mass of any planet to be $M\, sin(i)<$ 53 M$_\oplus$ in the Habitable Zone (1.2 AU). Examination of their Figure 6, which includes their estimates for the effects of non-Gaussian noise sources (``red noise") suggests a limit between 50 and 100 \mearth. This limit applies to the near edge-on, 79$^o$, orientation of the \acen\ AB system where dynamical studies indicate the presence of a stable zone $\lesssim$ 2.8 AU (or 2\farcs1) around \acen\ A despite the presence of \acen\ B \citep{Quarles2016,Holman1999,Quarles2018a,Quarles2018b}.

Due to the proximity of the \acen\ system, observations by \citet{Liseau2015,Liseau2016} with the Atacama Large Millimeter/submillimeter Array (ALMA) detected both stellar photospheres, allowing a unique opportunity to measure astrometric motion with ALMA.   Recently, the first radio astrometric planetary detection was announced by \citet{Curiel2020} with the discovery of a planet around the M dwarf TVLM 513–46546 using the Very Long Baseline Array.

In this paper, we present results from an ALMA observing program designed both to test the astrometric capabilities of ALMA and to make initial measurements of the \acen\ AB system.  Section \ref{sec:obs} describes our ALMA observations, data reduction, astrometric measurements and the contribution of stellar activity to the astrometric noise.  Section \ref{sec:results} includes our determination of the orbital and physical parameters for \acen\ A and B, combining our new data with previous observations, and a discussion of future prospects for planet detection.  Our conclusions are given in Section \ref{sec:conclusions} and Appendix A provides the ephemeris of \acen\ A and B.

\section{Astrometry with ALMA}
\label{sec:obs}

\subsection{Observations}

The properties of \acen\ A and B are well known from previous observations.  At sub-mm wavelengths, the stellar fluxes range from 10 to 100 mJy \citep{Liseau2016}, the angular sizes are 8.5 and 6.0 mas \citep{Kervella2017a}, the orbital separation ranges from 2 to 21\arcsec and the short term motion on the sky is 0.8 mas per hour.  From these parameters and looking to balance the SNR, the angular resolution and the field of view, we selected to observe in band 7.
As the astrometric precision scales directly with the angular resolution, observations when ALMA is in configurations with maximum baselines of 2 km or larger are most desirable, which limits the time sampling. Hence, observations were requested in three configurations, one in Fall 2018 and two in Summer 2019.     We obtained a series of ALMA observations of \acen\ in Cycle 6 between 2018 October and 2019 August 
at band 7, using the nominal continuum setup of four 1.9 GHz spectral windows at frequencies 336.5, 338.5, 348.5 and 350.5 GHz (Table \ref{tab:obs}).  To keep both sources well
within the primary beam, we placed the phase center for each 70-min observation at the center of
mass of \acen\ A and B. 

To assist in the determination of systematics in the astrometric measurements, we designed the observations   with two levels of repetition.  Within a configuration, we requested two observation blocks  separated by up to 14 days.  Within an observation block, we executed two independently calibrated observations, with little or no time gap in between them.  In 2018 October, we obtained only a single observational block, and one of the two observations within this block has issues (see Sec \ref{AbsA}) and is not included in the astrometric data results.  Thus, our final observation set includes nine calibrated observations, from five blocks:  one block
in configuration 6 (max baseline 2.5 km), two blocks in
configuration 8 (max baseline 8 km) and two blocks in configuration 7
(max baseline 3.6 km).

Each observation block consisted of the following: (1) The standard
on-line calibrations: pointing, delay offset, and system temperature
of each antenna; (2) a 2-min scan at the beginning of each observation
of the bright quasar, J1427-4206, with a known flux density; (3) the
main science target portion of the observation of about 55 observing
cycles, each with an 18-sec scan on the phase calibrator J1452-6502,
alternating with a 50-sec scan on \acen\ ; (4) within the science portion, 
five of the \acen\ target integrations were replaced by a check
source, J1424-6807.

\begin{deluxetable}{llllll}
\tabletypesize{\scriptsize}
\tablewidth{0pt}
\tablecaption{ALMA Observation Log \label{tab:obs}
}
\tablehead{
\colhead{Observation Pair} & \colhead{Maximum} & \colhead{Beam} &\colhead{rms} &\colhead{A int. flux\tablenotemark{a}} &\colhead{B int. flux\tablenotemark{a}}\\
\colhead{Dates} & \colhead{Baseline (km)} & \colhead{(mas)}& \colhead{($\mu$Jy)} &\colhead{(mJy)} & \colhead{(mJy)}    
}
\startdata
2018 Oct 14\tablenotemark{b} & 2.5 & 160 x 83 & 38 & 20.77$\pm$0.15 & 9.07$\pm$0.15 \\
2019 July 15-16 & 8.5 & 40 x 25 & 27 & 22.34$\pm$0.058 & 9.71$\pm$0.055\\
2019 July 19-20 & 8.5 & 35 x 21 & 30 & 21.19$\pm$0.081 & 9.83$\pm$0.081\\
2019 Aug 12-13 & 3.6 & 68 x 57 & 49 & 22.60$\pm$0.066 & 10.59$\pm$0.065\\
2019 Aug 25-26 & 3.6 & 69 x 55 & 50 & 26.08$\pm$0.051 & 11.91$\pm$0.051\\
\enddata
\tablenotetext{a}{Integrated flux uncertainty from fit; does not include absolute flux error}
\tablenotetext{b}{Only one observation block observed in this configuration}
\end{deluxetable}

\subsection{Data Reduction}

We used the standard ALMA data calibration for all  observations.
The amplitude calibration for each antenna/spectral-window channel  
was determined from the 2-min scan of J1427-4206 and
applied to the rest of the observation.  The measured antenna/spectral-window scale factor
is constant within 5\% over the $\sim$hour observation duration.  The phase
difference between each antenna/spectral-window
was essentially constant, and also obtained from the 2-min scan and
applied to the entire observation.  Thus, the eight data channels (four
spectral windows times two polarizations) could be combined to give
high SNR for the imaging, self-calibration and position determinations.
The phase calibration for \acen\ and the check source was obtained
in the normal phase referencing method from the antenna-based phase
from each of the fifty 18-sec scans on J1452-6502. \acen\  and
the check source field were then calibrated using the average of the
antenna-based phase of J1452-6502 just before and after the relevant
scan.

To provide an independent check of the absolute accuracy of these
observations, we included five 20-sec scans of the quasar J1424-6807 near \acen\
in each epoch as a check source.  This quasar is in the International VLBI Service for Geodesy and Astrometry list, and has a celestial position accuracy $<$0.5 mas.  Except for the integration time, which was
about 3.5\% of that of \acen, the identical check source observation and analysis methods 
provide an independent
confirmation of the absolute precision of the ALMA results.

From this calibrated data, we used the {\it tclean} program from the {\it casa} software package \citep{McMullin2007} to produce images for each of the calibrated observations, using the nominal data weighting (weighting=briggs, robust=0).  The image
size of 2048$\times$2048 pixels, with grid separation 5 mas, covered nearly all of
the ALMA primary beam.  The clean deconvolution algorithm 
removed the point spread function associated with the spatial frequency (uv) coverage
of the observation.  The image resolution varied from 20 mas to 150
mas, depending on the configuration.  We derived all astrometric positions from the images;  
we did not use uv data
analysis to obtain the source positions.  Using the stellar radii and parallax from \cite{Kervella2017a}, the angular diameters of \acen\ A and B are 8.5 and 6.0 mas respectively.  Thus, the individual photospheres are not resolved in these observations.
For the J1424-6807 check source, we analyzed an image size of 512$\times$512 pixels.

The flux for \acen\ A and B for each epoch is given in Table \ref{tab:obs}.  The integrated flux was determined using the {\it casa} routine {\it imfit} to fit a Gaussian to each star.  The uncertainty reported is the fitting uncertainty and does not include the absolute flux error.  As the focus of these observations was astrometry, no primary flux calibration standards were observed, thus the absolute calibration uncertainty is 10-15$\%$.  Within this absolute uncertainty, our fluxes agree with those of \cite{Liseau2016}.

\subsubsection{Absolute Astrometry} 
\label{AbsA}

The images made for each one-hour observation of \acen\ A and B are
on the VLBI J2000.0 International Celestial Reference System (ICRS) because the phase reference
calibrator, J1452-6502, and other interferometric parameters used for
these ALMA calibrations and reductions are on this frame associated with
very distant quasars.  The measured positions from each one-hour observation of \acen\ A and
B are shown in Table \ref{tab:absastrom}, as measured using the \emph{casa} routine \emph{jmfit}.  The positional change of each science target star due to proper motion and parallax is less than 1 mas in one hour. Each hour-long observation
can be combined into an average position with no significant loss of
precision in the absolute position.

\begin{deluxetable}{lllllll}
\tabletypesize{\scriptsize}
\tablewidth{0pt}
\tablecaption{Measured positions in IRCS \label{tab:absastrom}
}
\tablehead{
\colhead{Date} & \colhead{Start Time} & \colhead{Star} &\colhead{RA} &\colhead{Dec} &\colhead{$\sigma_{\rm RA}$ cos(Dec)\tablenotemark{a}} &\colhead{$\sigma_{\rm Dec}$ } \\
 & \colhead{UT hours} &  &\colhead{deg} &\colhead{deg} &\colhead{arcsec}  &\colhead{arcsec}
}
\startdata
14-Oct-2018 & 13:38:19.0 & A & 219.860763250 & -60.832171539 &  0.0039 &  0.0035 \\ 
& &B &219.859663542 & -60.830951061 &  0.0040 & 0.0039 \\
   15-Jul-2019& 23:14:41.3 &A& 219.858859933 &-60.832264944 &  0.0007 & 0.0010 \\ 
   && B& 219.857984904 &-60.830881649 &  0.0007 & 0.0010 \\ 
   16-Jul-2019&01:15:31.1 &A& 219.858854542 &-60.832262378 &  0.0019 & 0.0014 \\ 
   && B& 219.857979000 &-60.830879128 &  0.0018 & 0.0013 \\
   19-Jul-2019&23:31:14.5 &A& 219.858829571 &-60.832252293 &  0.0004 & 0.0007 \\ 
   && B& 219.857957646 &-60.830866720 &  0.0004 & 0.0007 \\ 
   20-Jul-2019 &01:02:29.9 &A& 219.858827083 &-60.832253354 &  0.0005 & 0.0005 \\ 
   && B & 219.857955246 &-60.830867652 &  0.0006 & 0.0007 \\ 
   12-Aug-2019 & 23:10:35.2 &A & 219.858667583 & -60.832185858 &  0.0068 & 0.0057 \\ 
   && B& 219.857814792 & -60.830785642 &  0.0071 & 0.0061 \\ 
   13-Aug-2019 & 00:44:57.6 & A & 219.858669208 & -60.832186997 &  0.0049 & 0.0037 \\ 
   && B & 219.857817458 & -60.830787197 &  0.0047 & 0.0035 \\ 
   25-Aug-2019 & 23:33:45.5 & A& 219.858615833& -60.832153722 &  0.0035 & 0.0020 \\ 
   && B & 219.857773917 & -60.830746378 &  0.0043 & 0.0024 \\ 
   26-Aug-2019 & 20:07:30.6 & A & 219.858613375 & -60.832152358 &  0.0016 & 0.0016 \\ 
   && B & 219.857772625 & -60.830744378 &  0.0018 & 0.0018 \\  
\enddata
\tablenotetext{a}{rms values given in table are the internal measurement uncertainties.}
\end{deluxetable}

We obtain a further check of the experiment accuracy from
analysis of the image position and stability of the check source
J1424-6807.  For the four highest resolution observations with a beam
about 30 mas taken in 2019 July, the average offset of the check source between our measured
ALMA position and its International Celestial Reference System (ICRS)
position is about 2.5 mas.  This displacement varied somewhat
with the quality and variability of the atmospheric phase stability
over each observation.  We note that the angular separation between the check source and the phase calibrator is 4.1 degrees, similar to the 4.5 degrees between \acen\ and the phase calibrator.

Extrapolating the absolute astrometric accuracy between the check source and \acen\ requires scaling with the peak SNR for each source.  The check source is roughly two times brighter than \acen\ B, but with only 3.5\% of the integration time, J1424-6807 will have a noise level $\sqrt(29)$ higher.  Thus the SNR ratio between \acen\ B and J1424-6807 is a factor of $\sim$3. In the absence of systematics, the best performance on \acen\ would scale to $\sim$0.8 mas.
As seen in Table \ref{tab:absastrom} the internal uncertainty for the best observation in 2019 July matches this performance.  The final absolute astrometric uncertainty will also include contributions from uncorrected phase noise in the \acen\ data. 
Based on variations in the absolute position of A and B using
different calibrations and array sizes, we use a conservative value of 3 mas
for the absolute astrometric uncertainty for the higher resolution configurations (2019 July and August) and 6 mas for the lower resolution (2018 October) data when
fitting the orbital parameters in Section \ref{sec:orbital}.  Future ALMA observations with good phase calibration on long baselines could provide
improved absolute astrometric accuracy of $<1$ mas, with internal
uncertainties of 0.1 to 0.2 mas in Band 7.

\begin{deluxetable}{lllll}
\tabletypesize{\scriptsize}
\tablewidth{0pt}
\tablecaption{Absolute Position Difference of Star A for five Observation Pairs \label{tab:absdif}}
\tablehead{
\colhead{Avg Date} & \colhead{RA meas} &\colhead{RA err cos(Dec)} &\colhead{dec meas} &\colhead{dec err} \\
yr & \colhead{mas} & \colhead{mas}& \colhead{mas}& \colhead{mas} } 
\startdata
  2018.7853 & -6.6 & 3.3 & 28.2 &  3.0 \\
  2019.5371 &  6.1 & 1.0 & -6.1 &  1.0 \\
  2019.5480 &  2.3 & 0.4 &  2.5 &  0.6 \\
  2019.6138 & -0.7 & 1.4 &  3.1 &  1.0 \\
  2019.6505 & -2.7 & 2.2 &  3.7 &  1.3 \\
\enddata
\end{deluxetable}

An additional estimate of the absolute position accuracy of \acen\ A
and B can be obtained by the position difference between each of the
five observation pairs in the experiment.  Table \ref{tab:absdif} shows the RA and declination differences between the two position measurements in each pair, after correction for the proper motion.  
The $RA meas$ column gives the measured difference between the two observations
separated by approximately one hour, and the $RA err$ column gives the 
uncertainty extrapolated from the image fitting errors given in Table 2.
Similar entries are given for the declination. The measured differences  
are correlated with the tropospheric stability during the
observation. The measured position differences between the observation pair
positions are generally a factor of 3 larger than that expected from the
internal position error based on the fitting of the star on each image
alone.  A similar scaling is obtained for \acen\ B which is not
shown.  The one anomaly is for the observation on 2018 Oct 14 at
15.1h where both stars A and B declinations disagree between the
observation pairs on 2018 Oct 14 at 13.6.  This observation is discussed further below. 

For the second observation block in October 2018, both
\acen\ A and B had an unexpected offset of 20 mas, while the
check source position repeatability error was less than 4 mas.  We continue to
investigate this problem, and believe it was caused by an error in
the ALMA celestial position tracking system of \acen.  Hence, this observation is not used in
our analysis below.

\subsubsection{Differential Astrometry}

The angular separation between \acen\ A and B can be measured from the
image for each of the observation eopchs.  Although the image quality
and absolute position offsets may differ, the atmospheric distortions act nearly identically on \acen\ A and B.  Thus, their angular separation is nearly independent of the quality of the
image, but with an error consistent with the SNR of each star. An empirical relationship \citep{Monet2010} shows the dependence of astrometric precision, $\sigma$, on the  Full Width at Half Maximum, FWHM,  of the beam and SNR of the signal: $\sigma\sim$ FWHM/(2$\times$SNR). Our results are consistent with this expectation.

\begin{figure}[h]
\includegraphics[height=7in]{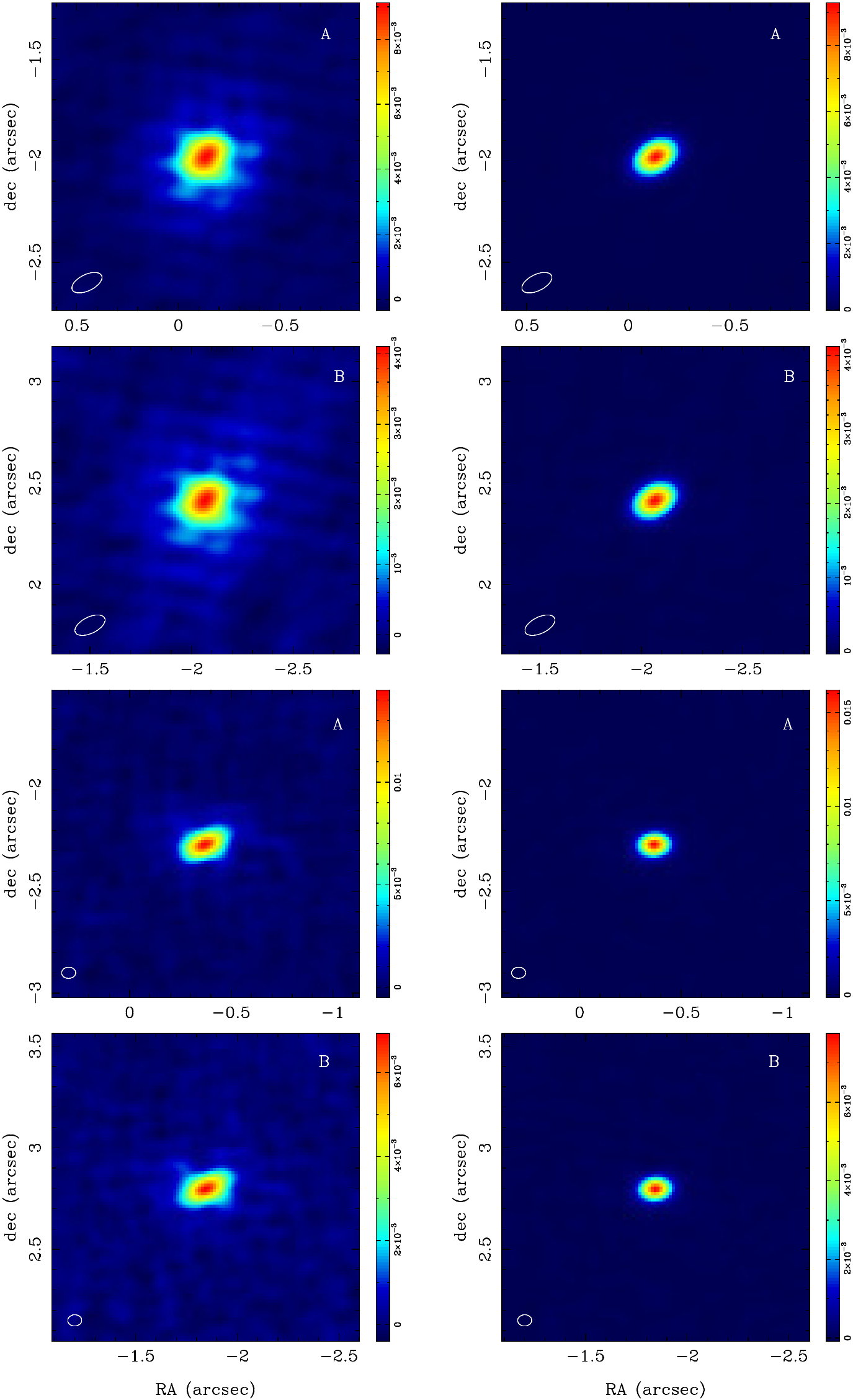}
\caption{Images of \acen\ A and B from 2018 October 14 (top 4 panels) and 2019 August 15 (bottom 4 panels).  Images on the left show standard phase calibration and images on the right show the corresponding data after phase self-calibration.  Each image is scaled independently to show the noise reduction after self-calibration.  The standard phase calibration images clearly show a larger size at the 10\% flux level due to smearing from uncorrected phase noise.
 \label{fig:ABimage}}
\end{figure}

As the radio emission from the two stars is sufficiently bright and compact,  we self-calibrate the \acen\ data in order to remove most of the residual tropospheric phase errors \citep{Sch80}. We used the self-calibration algorithm to improve the \acen\ image quality, and the following paragraph describes how the initial model for \acen\ assumed for self-calibration makes no significant difference on the A-B angular separation.
Figure \ref{fig:ABimage} shows an image of  \acen\ A and B before and after phase self-calibration, compared with an image obtained with only  normal phase referencing, for two observations with different spatial resolution. The improvement in the image quality is shown by the SNR, which increases by factor of $\sim$2 and by the significantly lower residuals.  Except for the emission from the two stars, we find no emission in the field of view greater than 0.3\% of the emission from \acen\ A.

Because we obtain significantly improved image quality  with self-calibration (Figure \ref{fig:ABimage}),  the \acen\ A to B separations measured from those images also have more precision and accuracy.  To check the impact of the self-calibration, the A to B separation was measured for 3 cases:  1) no self-calibration, 2) self-calibration for A only and 3) self-calibration for A and B. The separations for these three methods agreed to within the uncertainties and Table~\ref{tab:diffastrom} contains  \acen\ A to B separations measured from self-calibrated images for both sources. The angular separation and estimated error (using the casa
routine {\it jmfit}) are given in Table~\ref{tab:diffastrom}, measured
using self-calibrated method (3). The uncertainties in total separation are from the positional fitting and range from 0.3 to 0.6 mas. As absolute position information is not preserved during self-calibration, these data are only used for the A to B separation, not for the absolute positions given in Table \ref{tab:absastrom}.

\begin{deluxetable}{lllllll}
\tabletypesize{\scriptsize}
\tablewidth{0pt}
\tablecaption{\textbf{Relative position of $\alpha$\ Cen B with respect to A from ALMA.} \label{tab:diffastrom}
}
\tablehead{
\colhead{Date (UT)} & \colhead{Time (UT)} & \colhead{Decimal year} &\colhead{E-W separation cos(Dec)} &\colhead{$\sigma_{E-W}$ cos(Dec)} &\colhead{N-S separation} &\colhead{$\sigma_{N-S}$} \\
& \colhead{hours} & & \colhead{arcsec} & \colhead{arcsec} & \colhead{arcsec} & \colhead{arcsec} 
}
\startdata
14-Oct-2018 & 13.6 & 2018.78520 & -1.9310 & 0.0004 & -4.3952 & 0.0003 \\
15-Jul-2019 & 23.2 & 2019.53697 &    -1.5355 & 0.0003 & -4.9799 & 0.0003 \\
16-Jul-2019 & 01.3  & 2019.53720 &   -1.5361 & 0.0005 & -4.9802 & 0.0003 \\
19-Jul-2019 & 23.5 & 2019.54796 &    -1.5295 & 0.0003 & -4.9884 & 0.0003 \\
20-Jul-2019 & 01.0 & 2019.54813 &    -1.5297 & 0.0003 & -4.9884 & 0.0003 \\
12-Aug-2019 & 23.5 & 2019.61366 &   -1.4953 & 0.0003 & -5.0392 & 0.0003 \\
13-Aug-2019 & 01.0 & 2019.61385 &   -1.4945 & 0.0003 & -5.0390 & 0.0003 \\
25-Aug-2019 & 23.6 &  2019.64934 &   -1.4761 & 0.0005 & -5.0662 & 0.0003 \\
26-Aug-2019 & 20.1 & 2019.65168 &   -1.4751 & 0.0003 & -5.0684 & 0.0003 \\
\enddata
\end{deluxetable}

\subsection{Astrometric Noise due to Stellar Activity}

\cite{Dumusque2018} discusses the effects of stellar variability  in the context of PRV for planet detection, where changes in the shapes of spectral lines formed at different heights in stellar photospheres and chromospheres can adversely impact radial velocity precision at the $\sim 1$ m s$^{-1}$ level even for quiescent stars. A similar effect introduces spurious signals in astrometric observations. Time-variable spatial inhomogeneities on the surfaces of stars produce variations in their astrometric positions \citep{Makarov2009}. In the astrometric case, the causes of the variability include surface features such as sunspots and active regions, which may appear as fainter or brighter than the surrounding photosphere depending on wavelength. As spots rotate across and then behind the face of the star, the intensity-weighted centroid of the stellar disk will shift.
 
Appropriate to their age, both \acen\ stars are relatively quiescent with an activity index R$^\prime$HK= -5.15 and -4.97 for A and B, respectively \citep{Boro2018, Lisogorskyi2019}. Spectroscopic studies and  X-ray data \citep{Robrade2016} show \acen\ B to be modestly more active than \acen\ A, but both are similar in their activity levels to the Sun \citep{Mamajek2008}. \acen\ B is brighter at X-ray wavelengths \citep{Robrade2016} and shows a stronger chromospheric component \citep{Trigilio2018}. Both stars show cyclical variations in X-ray brightness over almost a 20 year span \citep{Robrade2016}.

\begin{figure}[h]
\includegraphics[height=0.5\textwidth]{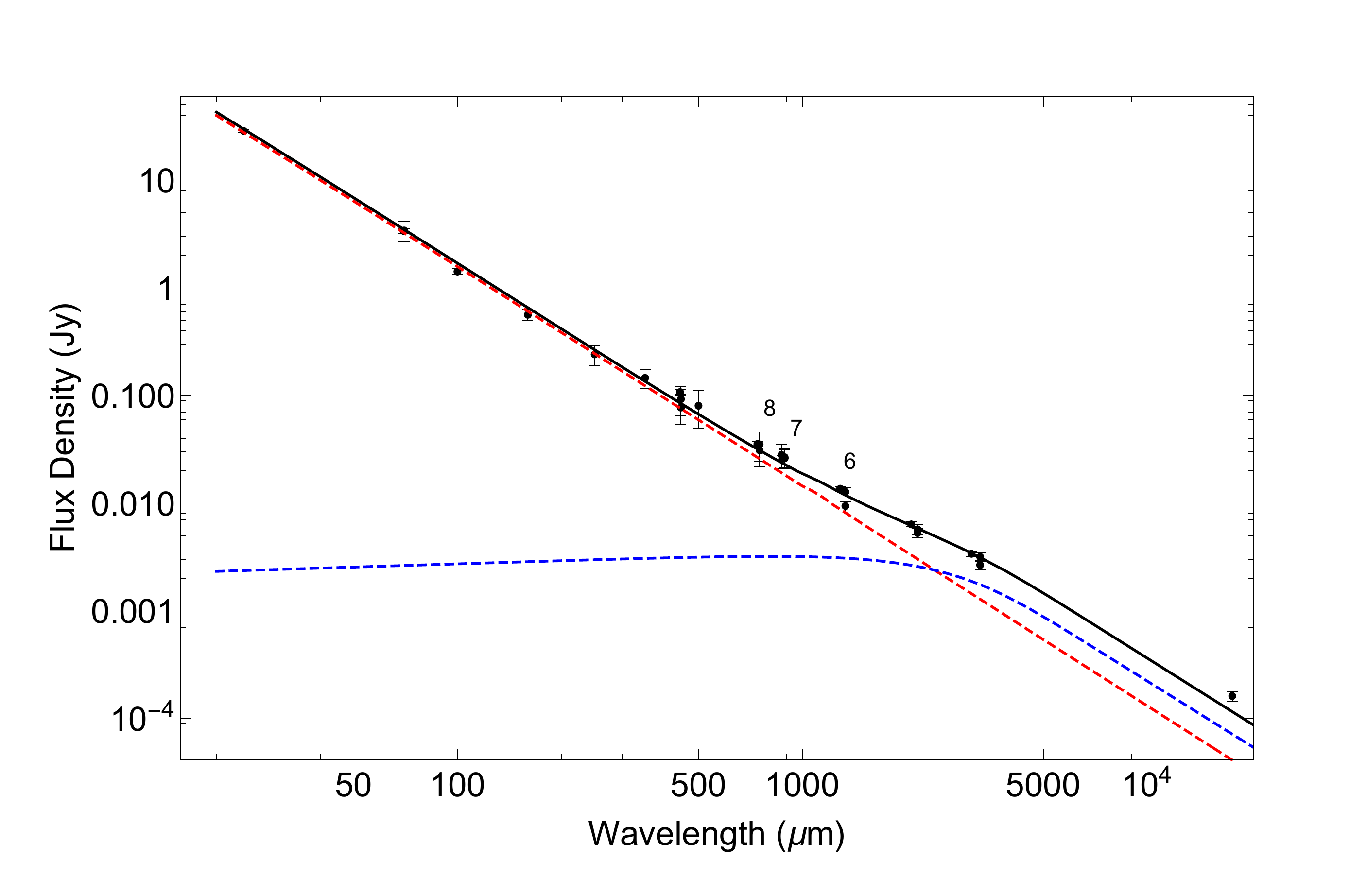}
\caption{A combination of Rayleigh-Jeans photospheric emission (red-dashed line) plus a free-free emission spectrum (blue-dashed line)  is fitted to photometry of \acen\ A. ALMA Bands 6, 7 and 8 are indicated. \label{photometry}}
\end{figure}

\acen\ A and B are strong sources at millimeter and sub-millimeter wavelengths. The spectral energy distribution in Figure~\ref{photometry} spans observations from the Spitzer, Herschel, ALMA and the Australia Telescope Compact Array (ATCA) observatories as tabulated in \cite{Liseau2013, Trigilio2018}  using ALMA values as updated with an improved calibration model  \citep{Liseau2019}.  To assess the possible impact of variability in the positions of the two stars we first must determine the dominant source of emission at our ALMA wavelength (Band 7; 343 Ghz or 874 $\mu$m). For \acen\ A, we fitted the absolute flux densities for a T$_{eff}=$5800 K B-T Settle model \citep{Baraffe2015} using R$_*=1.22R_\odot$ at 1.334 pc \citep{Kervella2017a} to the visible and NIR data listed in SIMBAD and \citet{Engels1981}. The fitted ``photospheric'' component is shown as the red-dashed curve in Fig~\ref{photometry} and has a normalization uncertainty of $9$\%\,$(1\sigma)$. To account for the excess obvious at longer wavelengths, we combine the photospheric emission with a simplified representation of free-free chromospheric emission following \citet{Mezger1967} with an optical depth given by:

$$\tau=3.28\times10^{-7}\left(\frac{T_{gas}}{10^4K}\right)^{-1.35}\nu^{-2.1}\times EM$$
\noindent and the corresponding flux density
$$F\nu=\frac{2 k_B T\nu^2}{c^2} \Omega (1-e^{-\tau}) $$

\noindent where $\tau$ is the free-free optical depth, $T_{gas}$ is the temperature of the ionized gas, $\nu$ the frequency, $EM$ the Emission Measure in pc cm$^{-6}$, $k_B$ is the Boltzmann constant, and $\Omega$ the solid angle of the star.

We fit the free-free emission from \acen\ A with a two parameter model, solving for gas temperature, T$_{gas}$, and emission measure, $EM$. The combination of photosphere (red-dashed line in Figure~\ref{photometry}) and free-free (blue-dashed line) emission provides a good fit to the data (solid black line) with $T_{gas}$=5100$\pm$650 K and $EM=3.0\pm 0.4\times10^{10}$ pc cm$^{-6}$. The derived T$_{gas}$ and EM (which corresponds to a 500-1000 km emitting region with an electron density of $\sim3\times10^{10}$ cm$^{-3}$) are comparable to the values in \citet{Trigilio2018}. 

The fit has a $\chi^2$ of 62 with 24 degrees of freedom. The high value of $\chi^2$ may be related to either variability in the multiple epochs of submillimeter data or to an underestimate of the absolute calibration of the data. We estimate the fractional contributions of the photosphere and chromosphere, $\Phi_p$ and $\Phi_c$, respectively, using the combined photospheric vs. chromospheric model  for Band 7. A Monte Carlo simulation varying $T_C$, $EM$ and the photospheric normalization, $1.02\pm0.09$, relative to their uncertainties gives a photospheric fraction of $\Phi_p=85\pm$1.3\% photospheric vs. $\Phi_c=1-\Phi_p=15\pm$1.4\% chromospheric. 
We performed a similar fit to the data for \acen ~B (T$_{eff}$=5300 K) and derived $\Phi_p=69\pm$7\% photospheric vs. $\Phi_c=31\pm$7\% chromospheric.

We are now in a position to assess the potential variability of the \acen\ stars. A detailed analysis \citep{Makarov2009} suggests that a quiescent star like the Sun would have a photospheric jitter at visible wavelengths of 1 $\mu$AU or 0.75 $\mu$arcsec at distance of 1.34 pc. Their model was based on an evolving population of sunspots on a rotating star with the starspots having an average temperature of T$_{spot}\sim$4000 K vs a photospheric temperature of T$_{phot}\sim$5800 K. At 0.55 $\mu$m the difference in surface brightness between these two regions would be $ {B_\nu(T_{spot}, 0.55\mu m)}/{B_\nu(T_{phot}, 0.55\mu m)}$=0.13 whereas in Band 7 (874 $\mu$m) the contrast would be much more muted, $ {B_\nu(T_{spot}, 874\mu m)}/{B_\nu(T_{phot}, 874\mu m)}$=0.69, where $B_\nu$ is the Planck function. Thus, a contrast-driven spatial jitter due to starspots in \acen\ A would be smaller by $0.69/0.13=5.3$, or only $\sigma_{phot}=0.14\, \mu$arcsec in the sub-millimeter. Unless \acen\ A or B is unusually active, which is not indicated by their R$^\prime$HK values, the purely photospheric contribution to astrometric jitter will be negligible.

\begin{figure}[h]
\plottwo{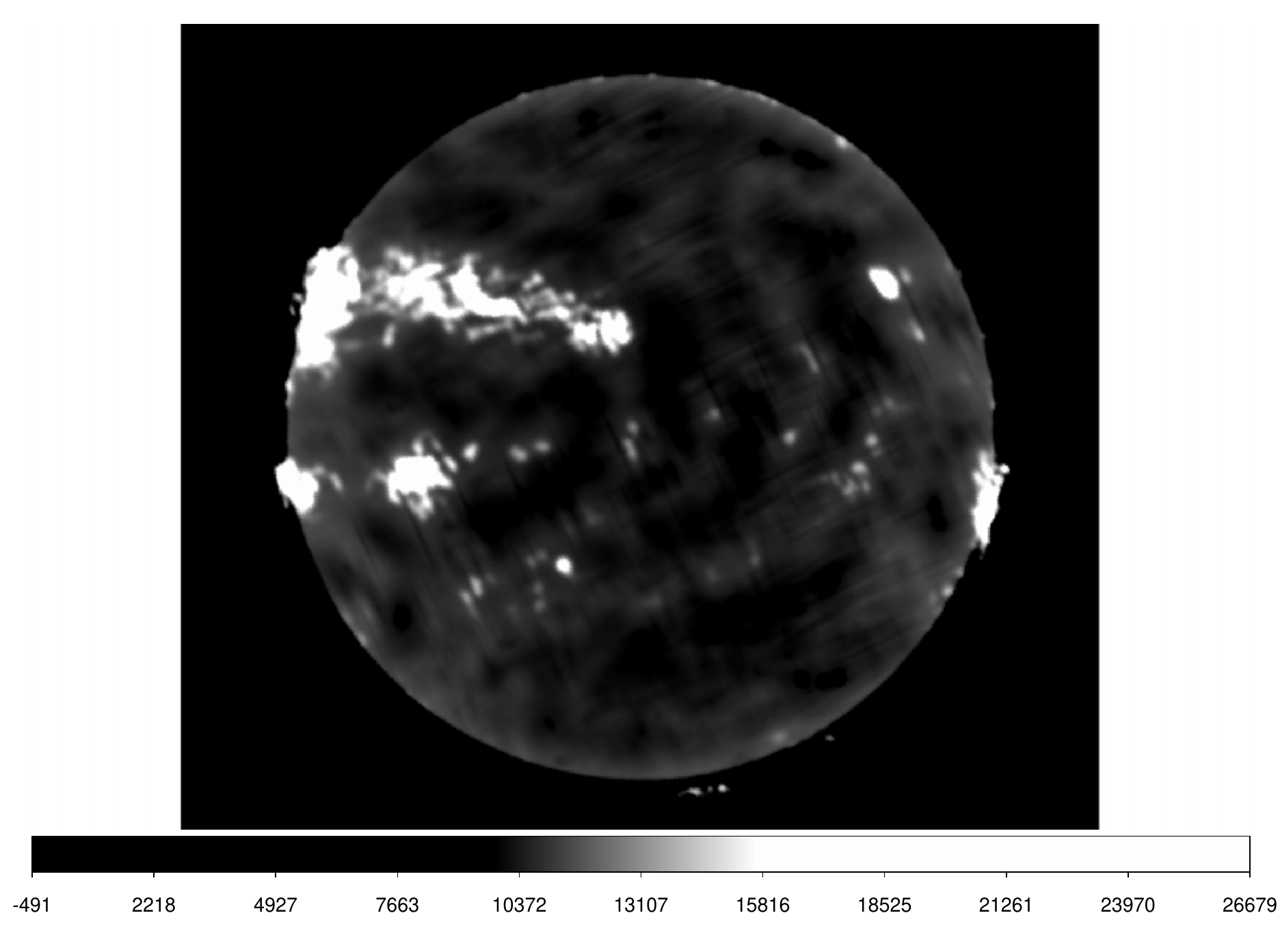}{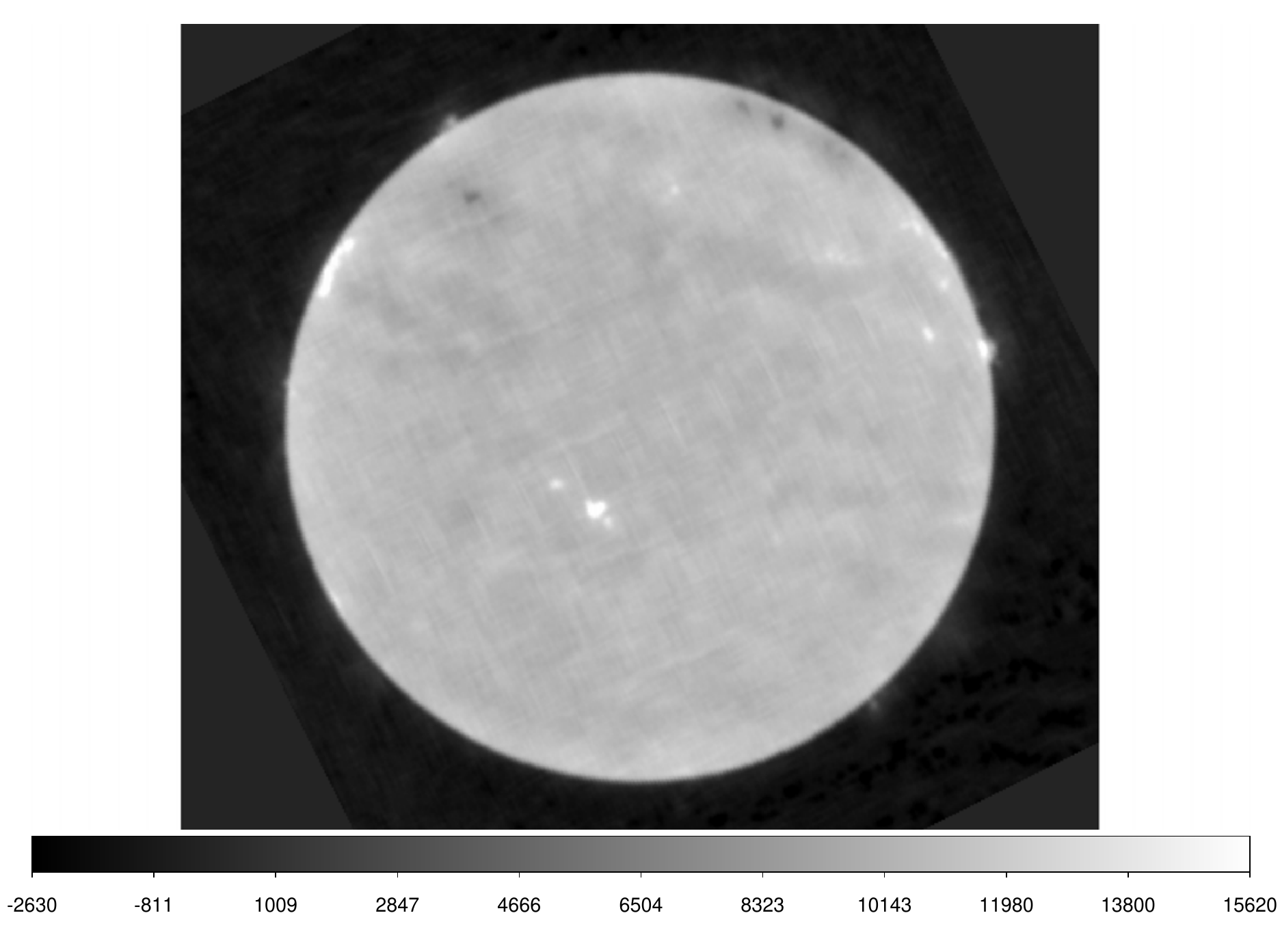}
\includegraphics[height=0.5\textwidth]{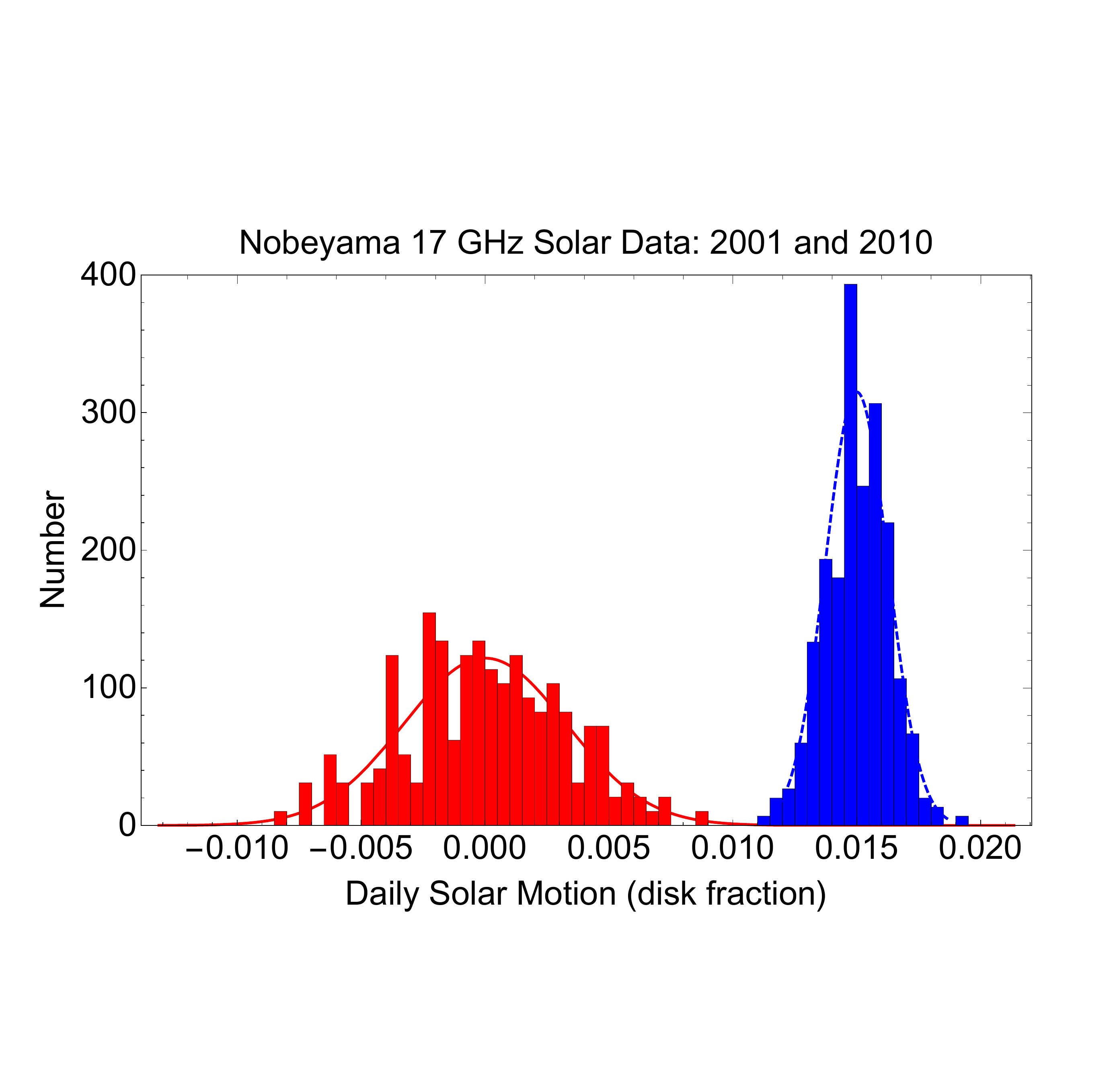}
\caption{Top-  Solar images from 2001 (left, Solar maximum) and 2010 (right, Solar Minimum) at 17 GHz from the Nobeyama telescope. Units are brightness temperature with quiescent regions having values around 10$^4$ K. Bottom- Histograms showing the number of days with varying amounts of offsets in the intensity-weighted centroid of the daily 17 Ghz maps  in 2001 (left, Solar Maximum in red) and 2010 (right, Solar Minimum in blue and displaced by 0.15 for clarity). Units are in fractions of the solar disk diameter. The 1$\sigma$ jitter in the centroid is $\sigma_\odot=0.36$\% (Active) and $\sigma_\odot=0.15$\% (Quiet). Gaussian distributions are over-plotted on the  histograms \label{solar}}
\end{figure}

We must still take into account any jitter in the free-free emission which comprises perhaps 15\% of the total emission from \acen\ A in Band 7. To assess this source of spatial variability we draw on a series of 17 GHz maps of the Sun obtained with 5\arcsec\ resolution taken daily by the Nobeyama Radio Interferometer \citep{Shibasaki2013}. Illustrative maps around Solar Maximum (2001) and Solar Minimum (2010) are shown in Figure~\ref{solar}a,b. We evaluated the shift between the intensity-weighted centroid and the nominal disk center by calculating the centroid of all chords across the solar disk in both the vertical and horizontal directions. We made this calculation for every day within 2001 (Solar Maximum) and 2010 (Solar Minimum) for which there were maps of suitable quality (244 in 2001 and 300 in 2010). Maps were subjected to visual examination to reject obvious artifacts. 

Figure~\ref{solar} shows the combined histogram of day-to-day shifts in the x and y centroids with a 1$\sigma$ standard deviation of 6\farcs46, or $\sigma_{\odot,active}=$0.36\% of the solar disk during 2001, and less than half that amount, 2\farcs64 or $\sigma_{\odot,quiet}=0.15$\% in 2010 (Table~\ref{jitter}). Multiplying these centroid shifts by the angular diameters of \acen\ A and B, $\theta_{A,B}$ \citep{Kervella2017a}, gives estimates of the jitter arising in the chromosphere, $\sigma_{c}=\sigma_\odot\theta$ expected from each stars in active or quiescent phases.

However, the influence of the chromospheric jitter term on the ALMA measurement is reduced by the contribution of the chromospheric emission to the total, $\Phi_c=32\pm13$\%. Propagating the Monte Carlo simulation of the photospheric vs. chromospheric faction into the jitter calculation yields the net amount of jitter expected for the two stars separately for the active and quiet states (Table~\ref{jitter}). These terms dominate the purely photospheric component due to starspots.

Finally, the astrometric jitter in the measurement of the {\textit separation between the two stars} is given by the root-mean-square sum of estimate for the two stars separately, $\sigma_{\rm tot}=\Phi_c \sqrt{\sigma_{c,\rm A}^2+\sigma_{c,\rm B}^2}$, and varies between 15 and 6 $\mu$arcsec in the active and quiet sun cases. This value represents only an order of magnitude estimate but suggests that for the purposes of astrometrically detecting planets significantly more massive than the Earth, stellar jitter terms are likely to be negligible in comparison with instrumental terms.

\begin{deluxetable}{lcc}
\tablecaption{Estimates of Astrometric Jitter\label{jitter}}
\tablehead{\colhead{Quantity} & \colhead{\acen\ A}& \colhead{\acen\ B}}
\startdata
Active Solar Jitter (Disk Fraction), $\sigma_{\odot,active}$& \multicolumn{2}{c}{0.36\%}\\
Quiet Solar Jitter (Disk Fraction),$\sigma_{\odot,quiet}$& \multicolumn{2}{c}{0.15\%}\\
Chromospheric Fraction, $\Phi_c$ &15$\pm$1.4\%&36$\pm7$\\\hline
Stellar Diameter, $\theta_{\rm A,B}$ (mas)\tablenotemark{a} & 8.5&6.0\\
Stellar jitter (Active, $\mu$arcsec),$\sigma_{\rm c,active}$& 4.6&8.5 \\
Stellar jitter (Quiet, $\mu$arcsec),$\sigma_{\rm c,Quiet}$& 1.9& 3.5\\
Total Jitter (Active, $\mu$arcsec),
$\sigma_{\rm tot,Active}$& \multicolumn{2}{c}{15$\pm$10}\\
Total Jitter (Quiet, $\mu$arcsec),$\sigma_{\rm tot,Quiet}$& \multicolumn{2}{c}{6$\pm$4}\\
\enddata
\tablenotetext{a}{\cite{Kervella2017a}}
\end{deluxetable}

\section {Results} 
\label{sec:results}

\subsection{Orbital and physical parameters of $\alpha$ Cen A and B}
\label{sec:orbital}

Our ALMA  observations have particular value for the determination of the orbit and proper motion of the $\alpha$\ Cen AB system. They are both the first precision absolute astrometric measurements of the positions of the two stars (referenced to the quasar frame) since {\it Hipparcos}, and the highest accuracy differential astrometric positions available to date. Due to the brightness and binarity of $\alpha$\ Cen AB, Gaia 
standard processing is unavailable and custom data reductions may be limited in absolute accuracy. 
However, the third component of the system, Proxima \citep{Kervella2017b}, is measured by Gaia with a high accuracy \citep{Kervella2020, Benedict2020}. Our analysis includes thousands of RV measurements of both components. The constraint provided by the requirement that astrometry and RV describe the same physical system improves the accuracy of our results.

\subsubsection{Astrometry Data}

To determine the orbital parameters of the system, we included archival \emph{differential} astrometry of the AB pair \citep{Hart01} starting in the year 1940 (one full 80\,yr orbit as of the year 2020).  Measurements older than 1940 do not  significantly improve the orbital parameters, due to their larger uncertainties and possible systematics. Our complete astrometric data set includes an {\it Hipparcos} measurement (from the data reprocessed by \citet{vanLeeuwen07}) combined with the new ALMA data points (Table~\ref{tab:diffastrom}) and these archival data. The complete set of differential measurements, including our new ALMA data, is given in Table \ref{tab:diffastromall}.

The only available high accuracy \emph{absolute} astrometric measurements of \acen\ A and B are the {\it Hipparcos} data and our new ALMA data points from 2018 and 2019.  Our complete sample of absolute astrometric measurements includes a total of 10 points; 9 from our ALMA data as given in Table \ref{tab:absastrom} and 1 from Hipparcos \citep{vanLeeuwen07} UT = 1991-04-02T13:30:00.000, RA$_{\rm A}=219.92040813\pm0.00307$, dec$_{\rm A}=-60.83514522\pm0.00246$, RA$_{\rm B}=219.91412460\pm0.01915$, dec$_{\rm B}=-60.83948046\pm0.01424$. We assign a systematic uncertainty of $\sigma = 3$\,mas to each absolute ALMA astrometric measurement, except for the lower resolution measurement from October 2018, which is assigned an uncertainty of 6 mas. These measurements determine the barycentric proper motion. Previous ALMA observations of alpha Cen from \citet{Liseau2015,Liseau2016} were designed for accurate flux measurements but not for accurate astrometric measurements and are not included in our analysis.  In particular, the angular resolution is a factor of several worse than our highest resolution and cross checks using a check source or between adjacent observations are not possible with those observations. We do, however, use these data as part of the differential astrometric data.

\begin{deluxetable}{llllll}
\tabletypesize{\scriptsize}
\tablewidth{0pt}
\tablecaption{\textbf{Relative astrometry of $\alpha$\ Cen B with respect to A.} \label{tab:diffastromall}
}
\tablehead{
\colhead{Date (UT)} & \colhead{Position angle} &\colhead{Pos. angle unc.} &\colhead{Separation} &\colhead{Separation unc.} &\colhead{Reference} \\
\colhead{decimal yrs} & \colhead{deg} & \colhead{deg} & \colhead{arcsec} & \colhead{arcsec}
}
\startdata
2019.6505 & 343.7729 & 0.0032 & 5.27869 & 0.00030 &this paper \\ 
2019.6481 & 343.7560 & 0.0053 & 5.27686 & 0.00032 &this paper \\
2019.6127 & 343.4803 & 0.0033 & 5.25595 & 0.00030 &this paper \\
2019.6125 & 343.4726 & 0.0033 & 5.25637 & 0.00030 &this paper \\
2019.5470 & 342.9519 & 0.0033 & 5.21767 & 0.00030 &this paper \\
2019.5468 & 342.9539 & 0.0033 & 5.21761 & 0.00031 &this paper \\
2019.5361 & 342.8581 & 0.0053 & 5.21172 & 0.00032 &this paper \\
2019.5358 & 342.8634 & 0.0033 & 5.21125 & 0.00030 &this paper \\
2018.7846 & 336.2542 & 0.0089 & 4.80434 & 0.00056 &this paper \\
2016.1893 &	305.19	&0.30	&4.013	&0.02 &	K16 \\
2015.3326	&293.30	&0.60	&4.020	&0.04	&K16/ALMA \\
2014.9574	&288.28	&0.70	&4.081	&0.05	&K16/ALMA \\
2014.5426	&282.91	&0.30	&4.208	&0.02	&K16/ALMA \\
2014.5126	&282.84	&0.30	&4.184	&0.02	&K16/ALMA \\
2014.2410	&279.20	&0.30	&4.330	&0.05	&An15 \\
2012.7100	&262.70	&0.40	&5.050	&0.05	&An14 \\
$\cdots$ \\
\enddata
\tablerefs{K16: \citet{Kervella2016}; K16/ALMA: \citet{Kervella2016} measurements of data from \citet{Liseau2015}; An15: \citet{Anton15}; An14:\citet{Anton14}}
\tablecomments{Table 6 is published in its entirety in the machine-readable format.
      A portion is shown here for guidance regarding its form and content.}
\end{deluxetable}

\subsubsection{Radial Velocity Data}

 We adopted radial velocity (RV) data from the large series of HARPS measurements of \acen\ A and B obtained since 2004,  publicly available from the ESO archive. We tested incorporating additional radial velocity measurements from other spectrographs (in particular CES and UVES). However, the shifts in the RV zero point between these instruments result in additional variables in the orbital fit which  degrade the quality and reliability. The extended time coverage provided by these instruments does not compensate for this drawback, and we therefore rely on the homogeneous set of HARPS measurements.

As noted by \citet{Lisogorskyi2019}, a significant fraction of the spectra collected for $\alpha$\ Cen A were processed using an incorrect cross-correlation mask of type K1V (corresponding in fact to $\alpha$\ Cen B). We adopted the $\alpha$\ Cen B RV from these authors, rejecting the points with the wrong masks from our sample. We also rejected the high cadence observations of 2013, known to be affected by a significant intra-night drift and possible systematics. The RV data of \acen\ A are from the standard ESO pipeline. We could not recompute the $\alpha$\ Cen A RV from the raw HARPS data, using a correct spectral-type mask and the standard ESO pipeline, because that pipeline is not publicly accessible. 

We corrected the RV shift caused by the HARPS fiber exchange in 2015, as recommended by \citet{Trifonov2020}. The applied shift is $\Delta V_A = +13.33$m s$^{-1}$ and $\Delta V_B = +11.53$m s$^{-1}$, which we subtract from the pipeline RVs for epochs later than $JD=2457163$.
We then filtered out the RV measurements whose uncertainty is larger than 10\,m\,s$^{-1}$, and cleaned the strongest outliers with a $5\sigma$ clipping with respect to the best fit preliminary solution. The latter step rejects only 0.9\% of the measurements of A and 1.0\% of B. Our final sample of RV measurements includes 5184 RV measurements for A and 12383 for B. A sample of the radial velocity data are shown in Table \ref{tab:radvel} and the complete table is available online.  Finally, we checked that adopting the \citet{Trifonov2020} RV shift does not result in a significant change of the final orbital parameters.

Although the statistical uncertainties of the HARPS RV measurements are extremely small (a few 10\,cm\,s$^{-1}$), the effective jitter is significantly larger due to stellar activity, crosstalk between the two stars or instrumental effects for these extremely bright targets. Therefore we added in quadrature a uniform uncertainty of 2\,m\,s$^{-1}$ to all  measurements.
Following \citet{Pourbaix2016} \citep[see also][]{Pourbaix2002,Kervella2016}, we subtracted a constant velocity shift of $V_B = 314$\,m\,s$^{-1}$ from the RV measurements of $\alpha$\ Cen B, thereby accounting for its relative gravitational redshift and convective shift with respect to $\alpha$\ Cen A, whose intrinsic shift is assumed to be zero.

\begin{deluxetable}{llll}
\tabletypesize{\scriptsize}
\tablewidth{0pt}
\tablecaption{\textbf{Radial velocity measurements for \acen\ A and B from the HARPS instrument} \label{tab:radvel}
}
\tablehead{
\colhead{MJD} & \colhead{Star} &\colhead{Radial Vel.} &\colhead{Rad. Vel. $\sigma$} \\
\colhead{days} & & \colhead{km sec$^{-1}$} & \colhead{km sec$^{-1}$} 
}
\startdata
53039.36288 & A & -22.73789  &  0.00201 \\
53039.36333 & A & -22.73742  &  0.00201 \\
53039.36376 & A & -22.73696  &  0.00202 \\
53139.06947 & A & -22.70210  &  0.00200 \\
53139.07017 & A & -22.70149  &  0.00200 \\
$\cdots$ \\
\enddata
\tablecomments{Table 7 is published in its entirety in the machine-readable format.
      A portion is shown here for guidance regarding its form and content.}
\end{deluxetable}

\subsubsection{Orbital and proper motion fit procedure}

To fit the orbital parameters and proper motion, we generally followed the same approach as \citet{Kervella2016}. We first determined the orbital parameters (Keplerian orbital elements, dynamical parallax, total mass of the system, and mass ratio) for epoch 2019.5 based only on the RV and differential astrometry data and a simple Keplerian two-body model.
In a second step, we adopted the dynamical parallax and mass ratio to determine the barycentric proper motion from the absolute Hipparcos and ALMA absolute astrometric positions of $\alpha$\ Cen A and B. The position of the barycenter of the system was thus computed for each of the Hipparcos and ALMA measurement epochs. The space velocity of the barycenter of the AB system was then determined from this set of barycenter positions through a linear fit.
For both steps, the determination of the best-fit parameters was obtained using a classical multi-parameter Levenberg-Marquardt least-squares fitting algorithm on the complete data set. It is based on the \texttt{scipy.optimize.leastsq} routine of the SciPy\footnote{\url{https://scipy.org}} library.

Among binary stars, $\alpha$\,Cen is peculiar due to its proximity to the Earth and fast motion across the sky. As a consequence, we implemented in our orbital and proper motion model a few second-order corrective terms that have an influence on the orbital parameters and proper motion estimates.
First, the very fast motion of $\alpha$\ Cen induces a significant change in perspective of its space velocity vector (that comprises the RV and tangential velocity components). This geometrical effect has been known for decades \citep[see, e.g.,][]{vandeKamp1977}, but to our knowledge it was not taken into account in recent computations of the orbital parameters of the system. It affects the barycentric RV and the tangential proper motion, both changing with time. The RV geometrically increases with an acceleration of $\dot{V} = +0.421$\,m\,s$^{-1}$yr$^{-1}$, while the tangential  velocity is conversely decreasing with time.

Another effect of the space motion of $\alpha$\ Cen is a perspective-induced change of the inclination of the orbit as seen from the Earth. This creates a time-dependent variation of the apparent relative position of B with respect to A. In addition, as the distance to the Earth is presently decreasing, the apparent separation of the two stars is increasing. At the {\it Hipparcos} epoch (1991.25) the perspective change results in a shift in relative position of B compared to epoch 2019.5 of approximately 30\,mas, which is not completely negligible compared to the error bars.

We also corrected the differential and absolute astrometric measurements for the light time propagation effect, based on our initial estimates of the barycentric RV. Due to the $\approx -22$\,km\,s$^{-1}$ approach velocity of the system, the light time aberration on its apparent position on the sky is evolving, compared to our 2019.5 reference epoch. In other words, as $\alpha$\,Cen was  farther to the Earth in the past, its apparent position was slightly more ``lagging behind'' on its trajectory compared to its true position in space than it is today.

For these different second order effects (perspective, light time), the changes are slow and can be considered linear over decades or even centuries. For this reason, we took them into account as a perturbation to our input data set. In practice, we computed an initial fit of the data without accounting for the perspective and light time effects. This resulted in an initial estimate of a mean value of the barycentric parallax, radial velocity and proper motion vector. Based on these results, we corrected the input differential and absolute astrometric positions as well as the RV values of the two stars to bring them to our reference epoch of 2019.5. This reference time was arbitrarily chosen to correspond to the majority of our ALMA measurements, in order to limit the correlations between the proper motion vector components. For these computations, we extensively used the \texttt{SkyCoord} class and celestial coordinate system transformation routines available in the  \texttt{Astropy}\footnote{\url{https://www.astropy.org}} library. We then computed a second round of orbital and proper motion fitting on these corrected observations, to determine the actual orbital parameters valid for our 2019.5 reference epoch. We checked that a third iteration of fitting did not result in a significant change of the derived parameters.

Finally, we emphasize that a complete astrometric solution  is critical to removing the  effects of the motions of  \acen A and B individually and as a system before looking for  differential motions due to a planet. The 30  year  temporal baseline from Hipparcos and ALMA and the high precision of the example 5-yr observing ALMA program discussed in \S \ref{planetdetection} would allow us to de-trend the effects of the 80-yr \acen AB orbit to identify the short period signatures of planets at the 10s of $\mu$as level in $\sim$ 1 yr orbit ($\S$\ref{prospects}). 

\subsubsection{Orbital parameters, masses and barycentric proper motion}

\begin{figure}[h]
\includegraphics[width=0.8\textwidth]{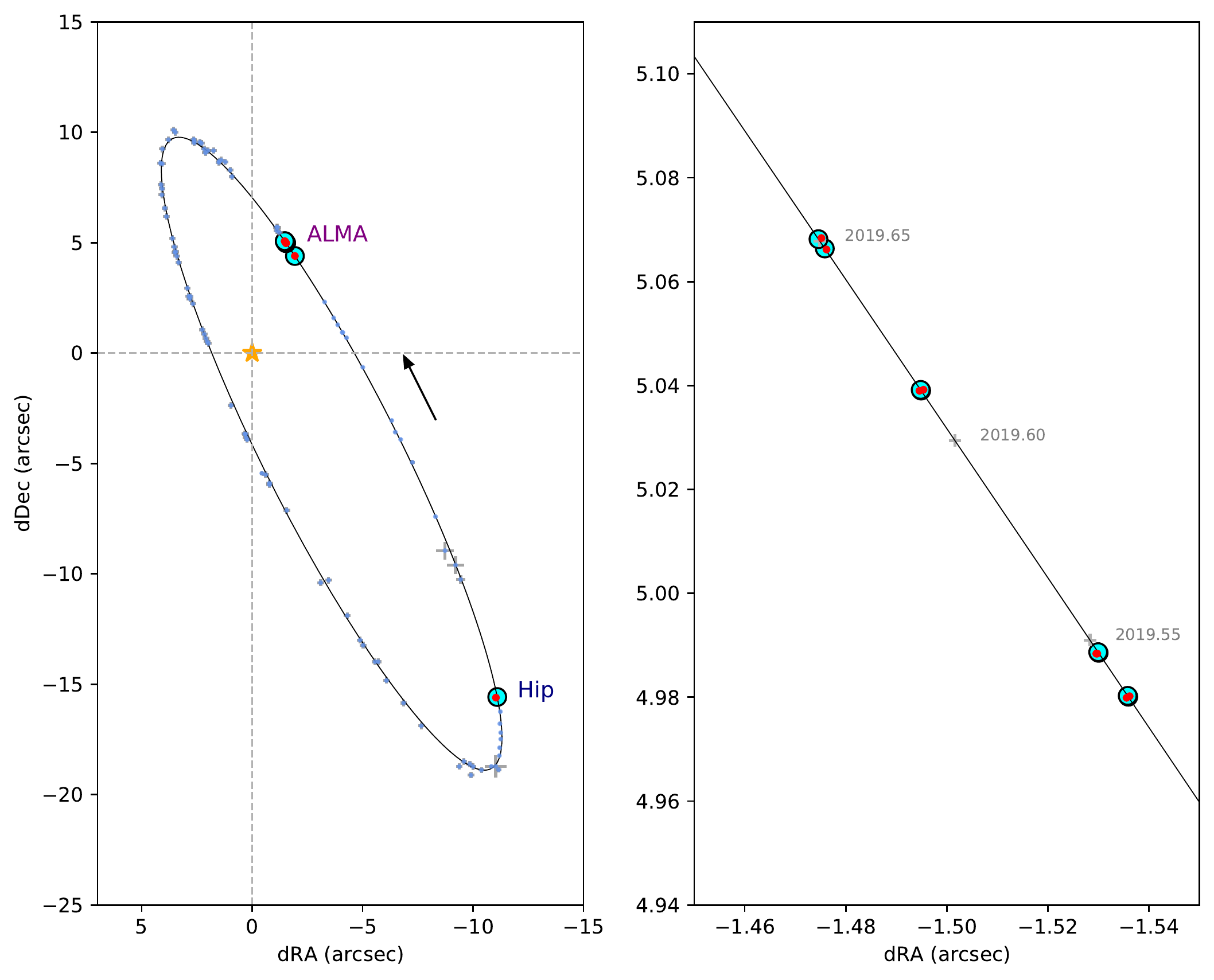}
\includegraphics[width=0.8\textwidth]{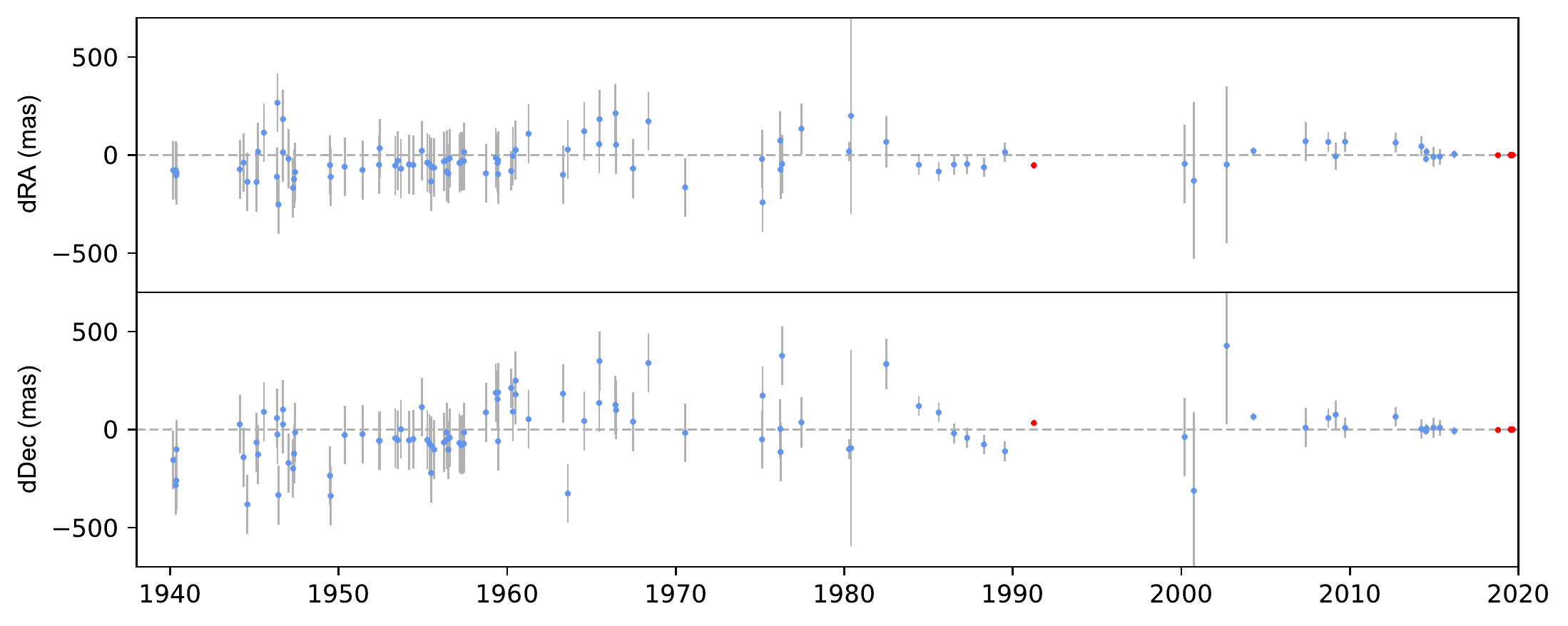}
\caption{Top left panel: Astrometric measurements and best fit orbit
  of \acen\ B relative to \acen\ A. The red dots represent the
  Hipparcos and ALMA data, while the light blue dots show the other
  measurements. Top right panel: enlargement of the 2019 ALMA
  measurements. In both panels, the cyan disks represent the model
  positions corresponding to the Hipparcos and ALMA points. Bottom
  panels: Residuals of the fit in right ascension and declination as a
  function of time. \label{orbit}}
\end{figure}

We list the best fit orbital parameters and masses of the $\alpha$\ Cen AB system  in Table~\ref{tab:orbital}, and present the corresponding orbital trajectory  in Figure~\ref{orbit}. We obtain a satisfactory reduced $\chi^2$ for the best fit orbit model of 1.2. The radii are computed adopting the limb darkened angular diameters measured by \citet{Kervella2017a}. The luminosities of A and B are computed adopting the bolometric flux by \citet{Boyajian2013} and $L_\odot = 3.828\ 10 ^{26}$\,W.

\begin{figure}[h]
\includegraphics[width=0.8\textwidth]{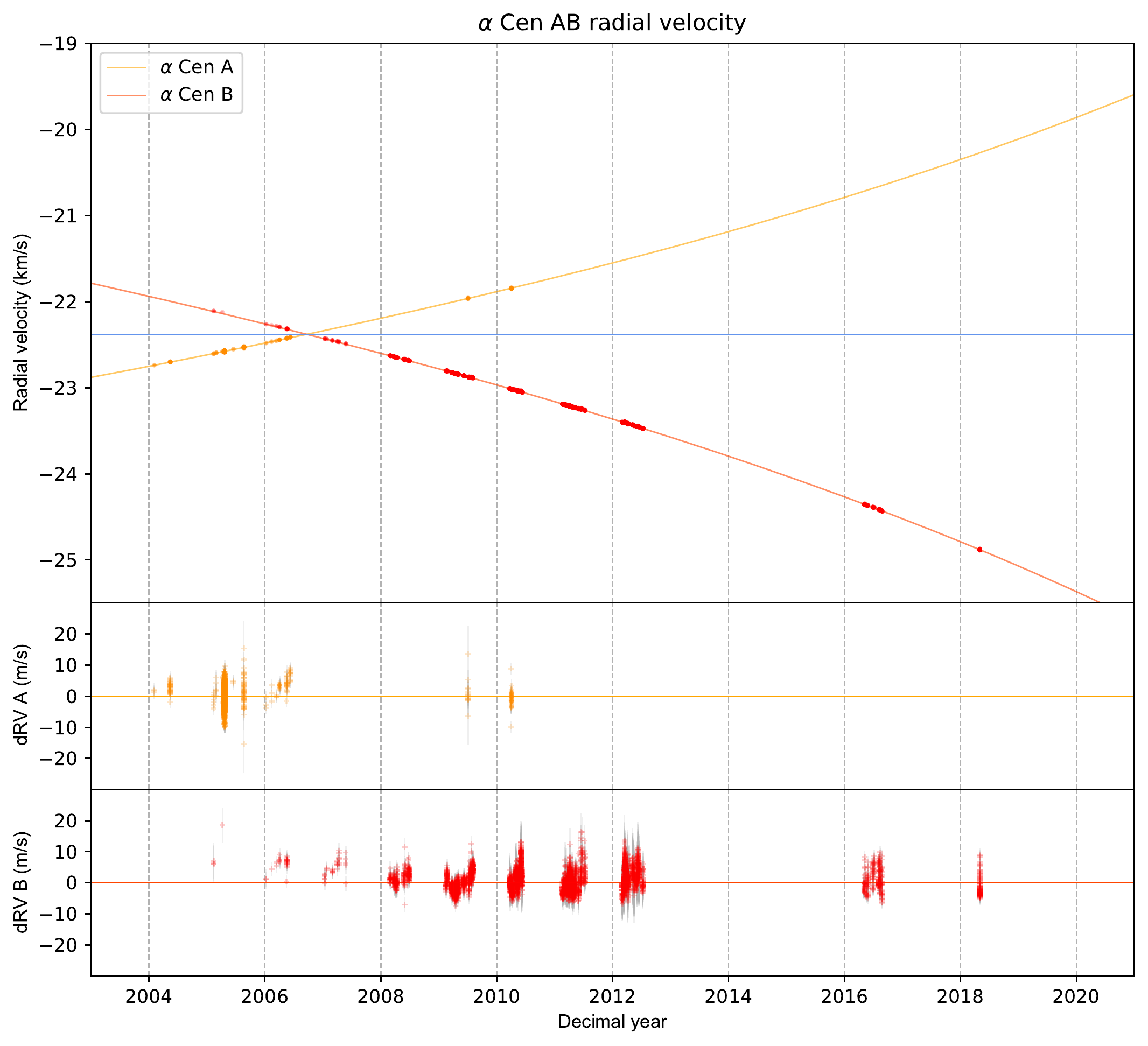}
\caption{HARPS radial velocity data compared to the best fit orbital model as a function of time (top panel) and residuals of the fit for \acen\ A (middle panel) and B (bottom panel). \label{RV_data}}
\end{figure}

\begin{figure}[h]

\includegraphics[width=9cm]{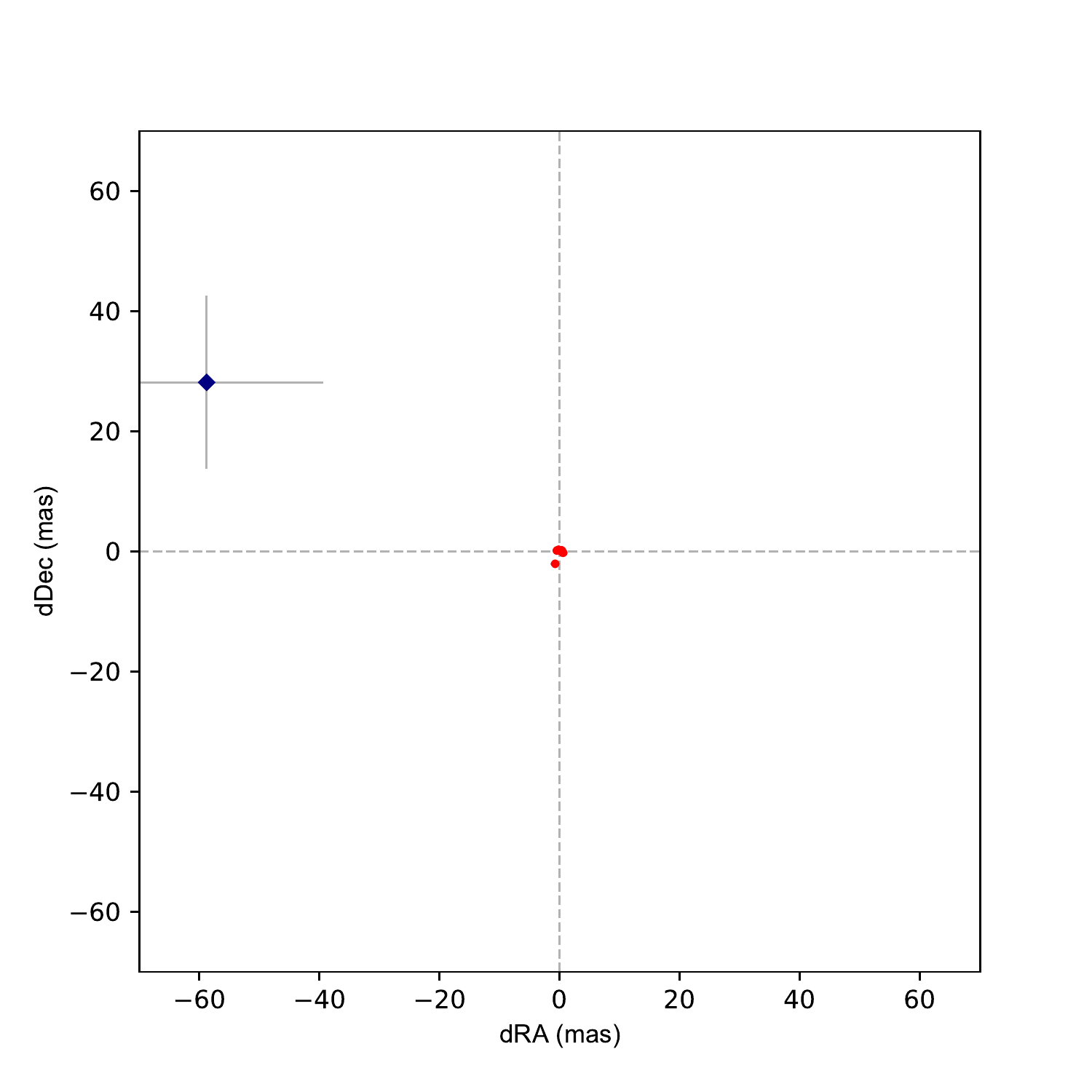}
\includegraphics[width=9cm]{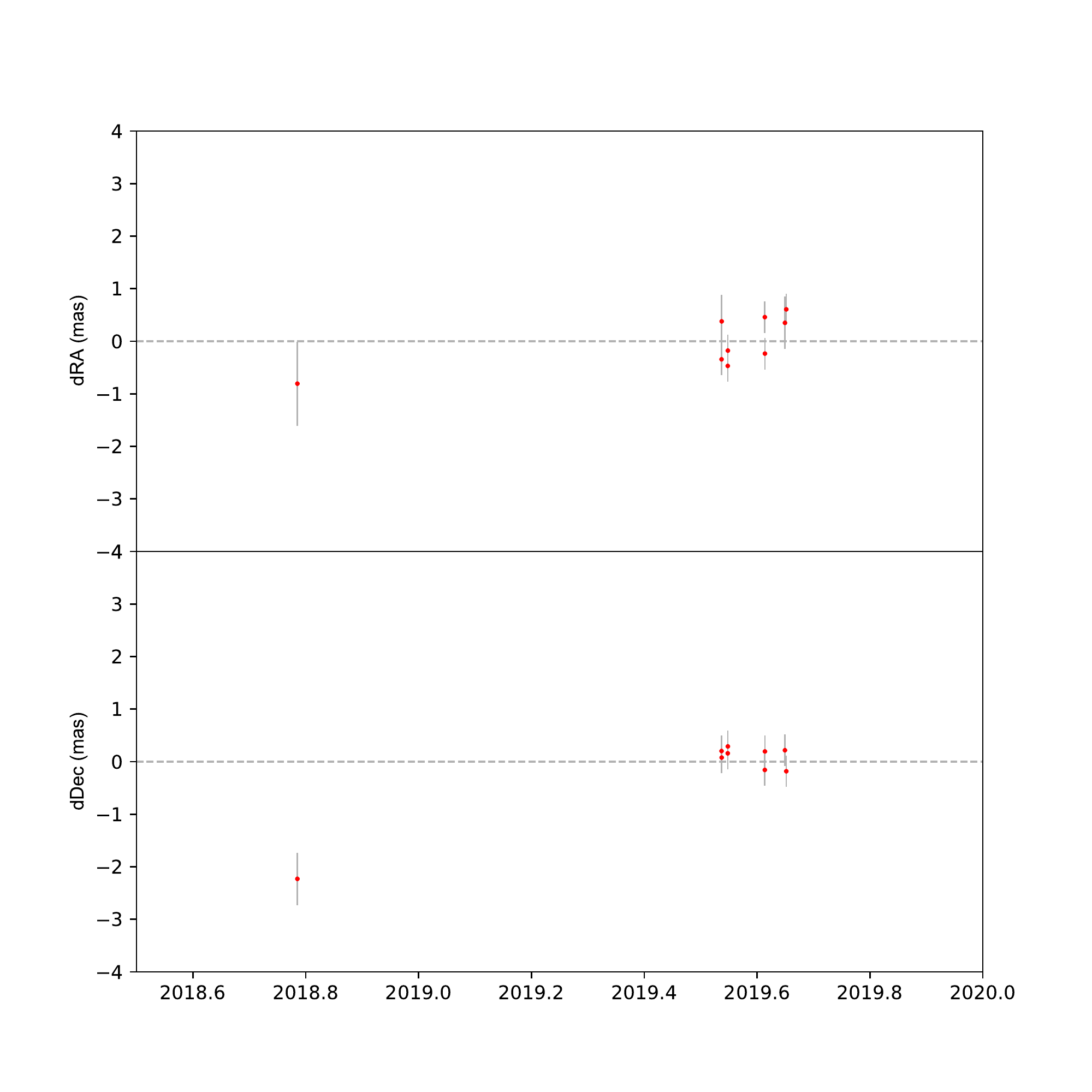}
\caption{Left panel: Differential position residuals of the ALMA (red) and Hipparcos (blue) measurements compared to the orbital fit of \acen\ B relative to \acen\ A on sky. Right panel: Right ascension and declination residuals of the ALMA differential A-B astrometric measurements with respect to the best fit orbit, as a function of time. \label{diffastrom_residuals}}
\end{figure}

We note  differences between the Table~\ref{tab:orbital} parallax and previous parallax determinations from \citet[hereafter PB16]{Pourbaix2016} and \citet[hereafter K16]{Kervella2016}. The main differences between these past analyses and the present work includes the selection of the RV data set and our new high accuracy ALMA differential astrometry. The radial velocity data used and the residuals of the fit are shown in Fig.~\ref{RV_data}. The acquisition of new RV measurements with HARPS, particularly of \acen\ A, would certainly improve the orbital phase coverage. We plot the residuals of the Hipparcos and ALMA absolute astrometric measurements   to the best-fit orbit  in Figure~\ref{diffastrom_residuals}.

\begin{deluxetable}{lclll}
\tabletypesize{\small}
\tablewidth{0pt}
\tablecaption{Orbital elements, masses, parallax, position and proper motion of \acen\ AB, for epoch 2019.5.   \label{tab:orbital}}
\tablehead{\colhead{Parameter} & & \colhead{Present\ work} &  \colhead{K16} & \colhead{PB16}}
\startdata
Semi-major axis (arcsec) & $a$ & $17.4930 \pm 0.0096$ & $17.592 \pm 0.013$ & $17.66 \pm 0.026$ \\
Inclination (deg) & $i$ &     $79.2430 \pm 0.0089$ & $79.320 \pm 0.011$ & $79.32 \pm 0.044$ \\
Arg.\,of periastron (deg) & $\omega$ &    $231.519 \pm 0.027$  & $232.006 \pm 0.051$ & $232.3 \pm 0.11$  \\
Long.\,of asc.\,node (deg) & $\Omega$      &    $205.073 \pm 0.025$ & $205.064 \pm 0.033$ & $204.85 \pm 0.084$ \\
Period (yr) & $P$   &     $79.762 \pm 0.019$ & $79.929 \pm 0.013$ & $79.91 \pm 0.013$ \\
Ref.\,epoch & $T_0$ &  $1955.564 \pm 0.015$ & $1955.604 \pm 0.013$ & $1955.66 \pm 0.014$ \\
Eccentricity & $e$ &     $0.51947 \pm 0.00015$ & $0.5208 \pm 0.0011$ & $0.524 \pm 0.0011$ \\
Barycentric RV (km\,s$^{-1}$) & $V_0$ &  $-22.3796 \pm 0.0020$ & $-22.3930 \pm 0.0043$ & $-22.390 \pm 0.0042$ \\
RV accel. (m\,s$^{-1}$\,yr$^{-1}$) & $\dot{V_0}$ & $+0.421$  & $-$ & $-$ \\
Parallax (mas) & $\varpi$ &  $750.81 \pm 0.38$ & $747.17 \pm 0.61$ & $743 \pm 1.3$ \\
Parallax var. ($\mu$as\,yr$^{-1}$)  & $\dot{\varpi}$ &   $+13.0$ & $-$ & $-$ \\
\hline
Mass fraction & $\frac{m_{\rm A}}{m_A+m_{\rm B}}$ & $0.54266 \pm 0.00011$ & $0.54116 \pm 0.00027$ & $0.5383 \pm 0.00044$ \\
Mass of A  $(M_\odot)$ & $m_{\rm A}$ &  $1.0788 \pm 0.0029$ & $1.1055 \pm 0.0039$ & $1.133 \pm 0.0050$ \\
Mass of B  $(M_\odot)$ & $m_{\rm B}$ &  $0.9092 \pm 0.0025$ & $0.9373 \pm 0.0033$ & $0.972 \pm 0.0045$ \\
\hline
Radius of A $(R_\odot)$ & $R_A$ & $1.2175 \pm 0.0055$ \\
Radius of B $(R_\odot)$ & $R_B$ & $0.8591 \pm 0.0036$ \\
Luminosity of A $(L_\odot)$ & $L_A$ & $1.5059 \pm 0.0019$ \\
Luminosity of B $(L_\odot)$ & $L_B$ & $0.4981 \pm 0.0007$ \\
\enddata
\end{deluxetable}

Table~\ref{tab:propermotion} lists  the position and proper motion of the barycenter of the \acen\ system for epoch 2019.5. The reduced $\chi^2$ of the best fit barycenter proper motion vector model is 1.6. For all the model fitting done for the present work, the uncertainties of the best-fit parameters are normalized to the data dispersion around the best-fit model. In practice, this means that the $1\sigma$ uncertainties of the derived parameters are scaled by a factor of $\approx \sqrt{1.6}$. Through this rescaling, we take into account in the uncertainties the residual dispersion of the data points with respect to the model, that may be induced by instrumental or astrophysical effects.

\begin{figure}[h]
\includegraphics[height=5.5cm]{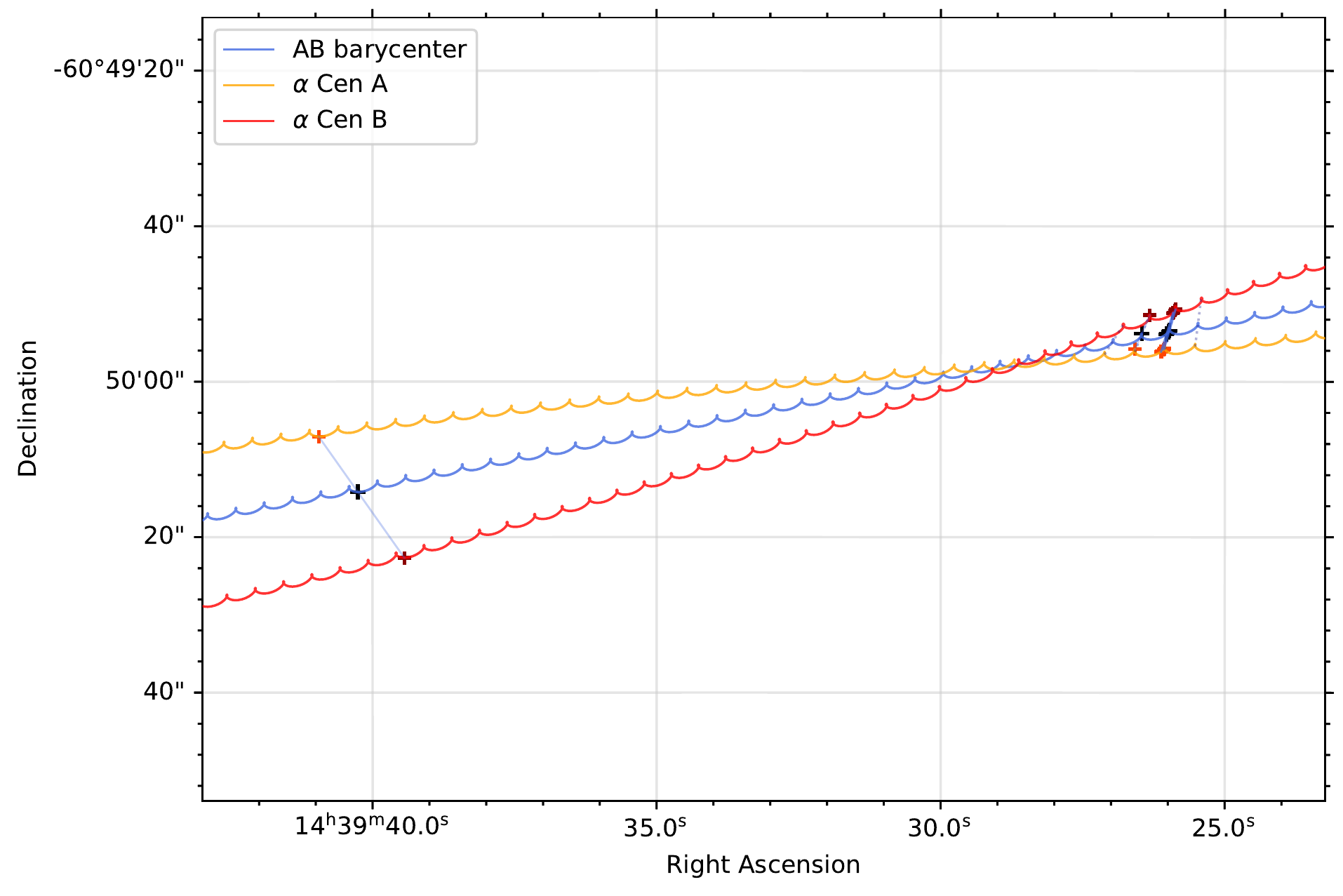}
\includegraphics[height=5.5cm]{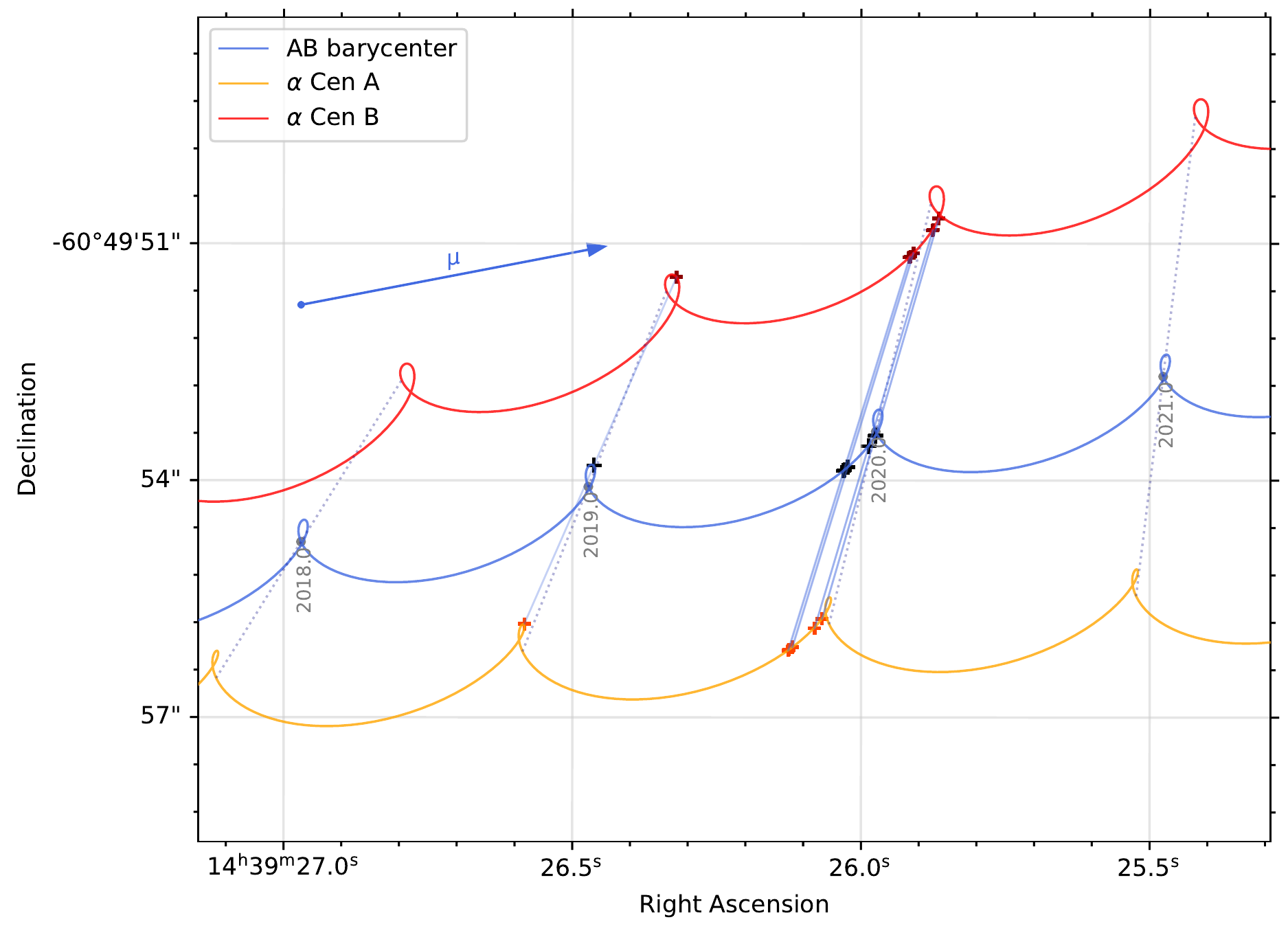}
\caption{Sky trajectory of \acen\ A, B and their barycenter, showing the Hipparcos and ALMA measurements (left panel) and an enlargement of the ALMA measurements (right panel). The  individual points are shown with '+' symbols. The $\mu$ vector represents the annual proper motion of the barycenter. \label{sky_trajectory}}
\end{figure}

\begin{figure}[h]
\includegraphics[width=9cm]{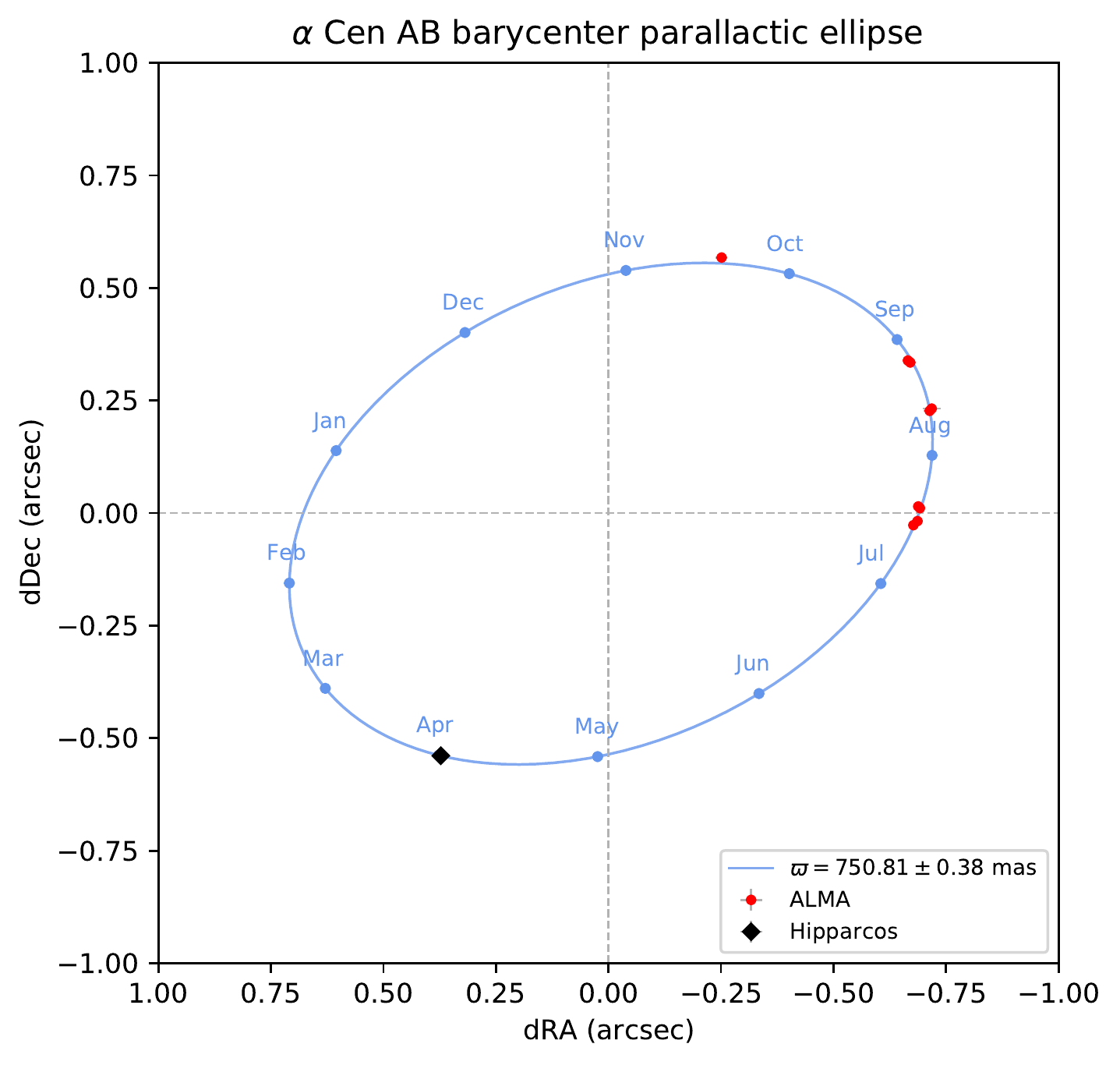}\\
\includegraphics[width=11cm]{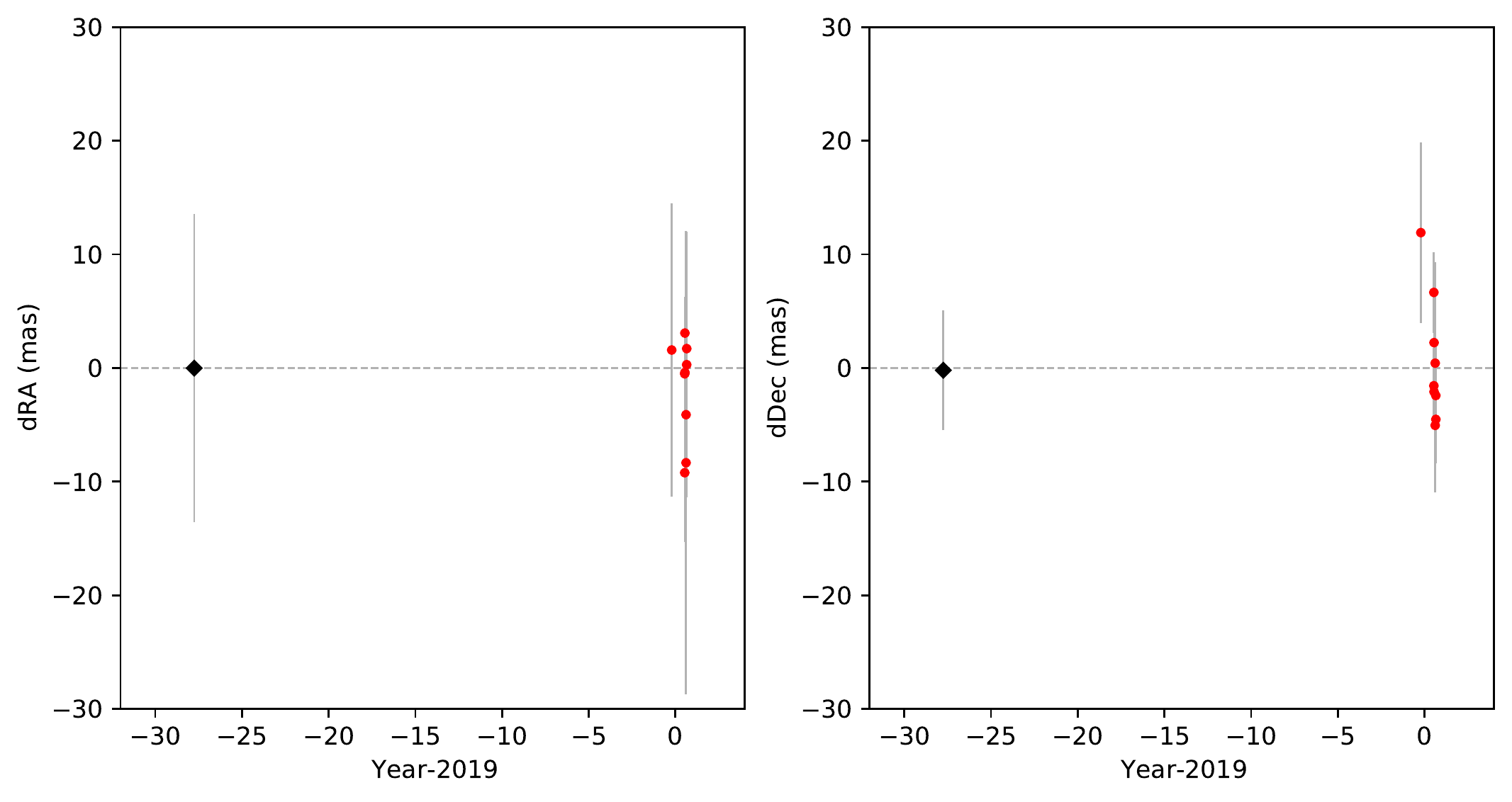}\\
\includegraphics[width=11cm]{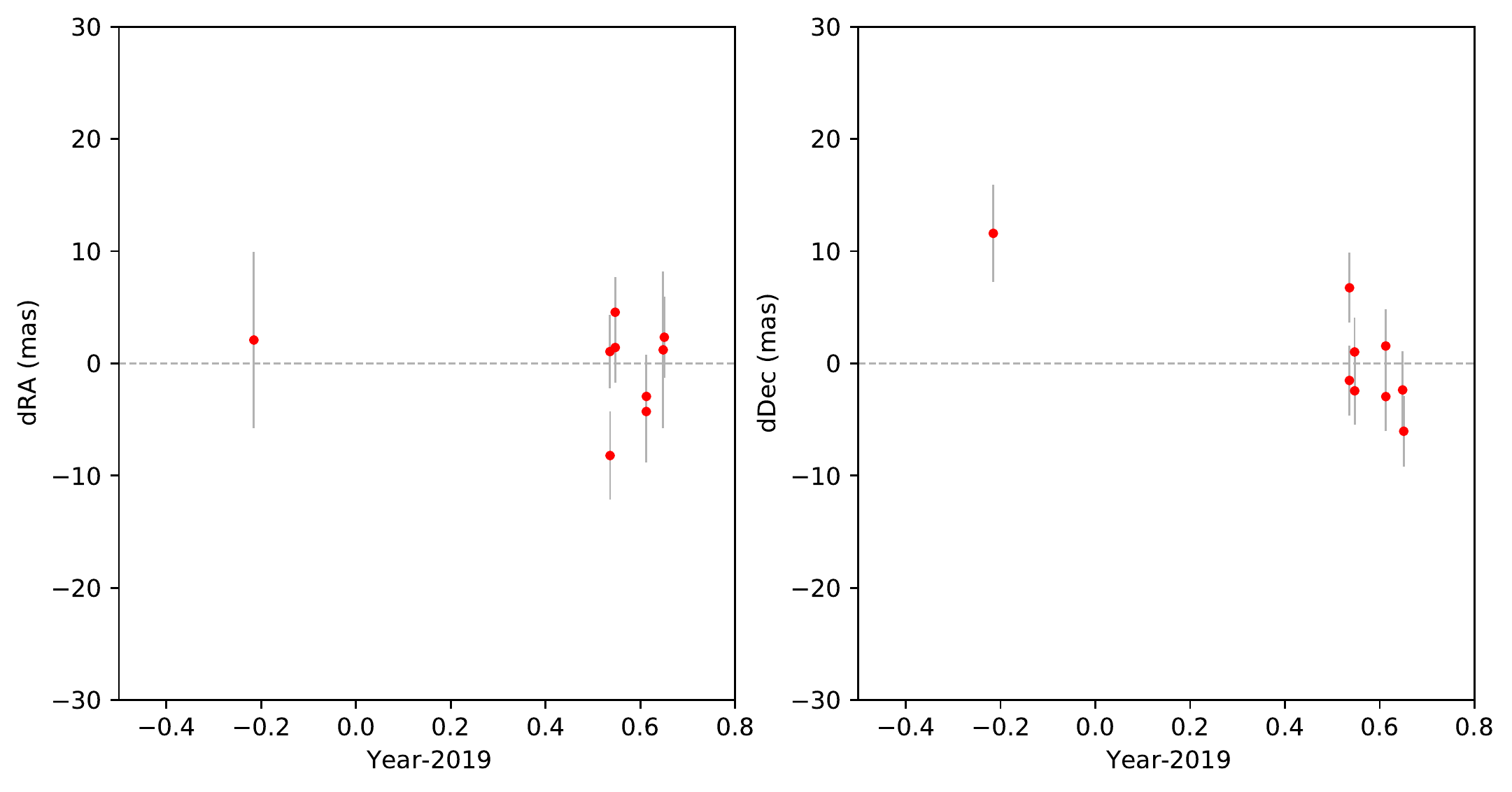}
\caption{Top panel: Positions of the barycenter of \acen\ AB measured with ALMA (red) and Hipparcos (black) on the parallactic ellipse of \acen\ (the proper motion has been subtracted). The  ellipse corresponds to the parallax derived from the AB orbital fit (this is not a fit). 
Middle panels: Residual positions of the barycenter of \acen\
  AB measured with Hipparcos (black) and ALMA (red points), compared
  to the best fit trajectory including proper motion and parallax. Bottom panels: Enlargement of the residuals of the ALMA measurements in 2018 and 2019 (red points).
\label{barycenter_ellipse}}
\end{figure}

Due to the perspective acceleration, the proper motion vector, together with the radial velocity, changes with time. The sky trajectory of \acen\ A, B, and their barycenter is shown in Fig.~\ref{sky_trajectory}. The apparent motion of A and B is the result of the combination of the proper motion of the barycenter, orbital motion of the two components, and the parallax.
Figure~\ref{barycenter_ellipse} displays the measured position of the AB barycenter compared to the parallactic ellipse derived from the orbital fit.
Although the agreement is satisfactory between the measurements and the ellipse, this is not a fit. It is possible to derive the trigonometric parallax by fitting the observed Hipparcos and ALMA positions of the barycenter. But, due to relatively poor phase coverage of the parallactic ellipse because of our limited number of absolute measurements, the accuracy of the best fit trigonometric parallax ($\varpi_\mathrm{trig} = 747.1 \pm 5.2$\ mas) is significantly worse than that of the orbital parallax ($\varpi_\mathrm{orb} =750.81 \pm 0.38$\,mas). Both values are, however, statistically consistent.

\begin{deluxetable}{lcll}
\tabletypesize{\small}
\tablewidth{0pt}
\tablecaption{Position and proper motion (PM) of the barycenter of \acen\ A and B. The parameters derived by \citet{Kervella2016} and \citet{Kervella2017b} are listed in the K16, K17 column for comparison. \label{tab:propermotion}}
\tablehead{\colhead{Parameter} & & \colhead{Present work (J2019.5)} &  \colhead{K16, K17 (J1991.25)}}
\startdata
RA (ICRS) & $\alpha_0$ & 14:39:26.1413 $\pm 1.50$ mas & 14:39:40.2068 $\pm 25$ mas \\
 & & $219.85892215 \pm 8.6\,10^{-7}$ deg & $-$ \\
Dec. (ICRS) & $\delta_0$ & $-60$:49:53.875 $\pm 1.17$ mas & $-60$:50:13.673 $\pm 19$ mas \\
 & & $-60.83163195 \pm 3.3\,10^{-7}$ deg & $-$ \\
PM RA (mas\,yr$^{-1}$)  & $\mu_\alpha$ &  $-3639.95 \pm 0.42$ & $-3619.9 \pm 3.9$ \\
PM Dec. (mas\,yr$^{-1}$)  & $\mu_\delta$ &  $+700.40 \pm 0.17$ & $+693.8 \pm 3.9$ \\
\enddata
\end{deluxetable}

\subsection{Prospects for Improved Stellar Properties\label{prospects}}

The \acen\ AB system is a touchstone for stellar astrophysics because these two stars straddle the Sun  in mass with similar age, activity level and metallicity. Because their physical properties can be determined with great precision due to their proximity to us, these stars are used  for validation of  stellar models and  to check the predictions of techniques such as asteroseismology, e.g. \citet{Joyce2018, Nsamba2019}. 

\acen\ is one of the Gaia mission benchmark systems selected as references for the determination of stellar parameters \citep{Heiter2015}. In advance of those anticipated, possibly more precise results, we discuss briefly how our new ALMA measurements improve the precision of our knowledge of \acen\ A and B.

The combination of asteroseismic frequencies and interferometric radius measurements has proven to powerfully constrain the internal structure of stars \citep{Cunha2007,Chaplin2013}. \acen\ A and B were among the first stars to have their oscillation spectra measured with high precision by \citet{Bouchy2001, Bouchy2002}, using radial velocity measurements with the CORALIE spectrograph.
The angular diameters of \acen\ A and B were first measured interferometrically with the VLTI/VINCI instrument by \citet{Kervella2003}, and later refined with the VLTI/PIONIER instrument \citep{Kervella2017a} at $\theta_\mathrm{LD}$(A) $= 8.502 \pm 0.038$\,mas and $\theta_\mathrm{LD}$(B) $= 5.999 \pm 0.025$\,mas.
The parallax determined in the present work ($\varpi = 750.81$\,mas, $d=1.3319 \pm 0.0007$\,pc) is larger by 0.49\% compared to the value previously determined by \citet{Kervella2016} ($\varpi = 747.17$\,mas).
As a result, we revise the linear radii of the two stars to $R_{\rm A} = 1.2175 \pm 0.0055\,R_\odot$ and $R_{\rm B} = 0.8591 \pm 0.0036\,R_\odot$, a change of approximately $-1\,\sigma$.
The masses and radii values are close to the theoretical values determined by \citet{Thevenin2002} and \citet{Yildiz2008} from the analysis of the asteroseismic frequencies of the two stars.

The determination of the age of stars is generally a complex enterprise \citep{Soderblom2010}. In spite of the similarity of the physical properties of \acen\ with the Sun and their well-determined masses, the estimate of their age \citep{Mamajek2008,Morel2018,Sahlholdt2019}, core properties \citep{deMeulenaer2010, Bazot2016}, and inner opacity \citep{Yildiz2011} remain a challenging task. For the evolutionary modeling of \acen\ A and B, the treatment of convection is problematical, and the mixing-length parameter remains relatively uncertain \citep[e.g.][]{Yildiz2007, Trampedach2014, Joyce2018, Spada2019}.

For \acen\ A and B, further progress on the modeling of structure and evolutionary status requires improvements to fundamental stellar parameters, both in higher precision and accuracy. \citet{Joyce2018}  give the age of \acen\ AB  as 5.3$\pm$0.3 Gyr based on masses which are $\sim 3\%$ higher than those presented here. New models using these more precise values may result in revised ages for this touchstone system. While the uncertainties in  the ages derived from the model fitting  due to uncertainties in mixing length, etc,  may remain the same,  the centroid of the age estimates  may shift significantly, especially as future ALMA observations reduce the parallax uncertainties by another factor of 2-3.

\subsection{Prospects for Planet Detection based on Differential Astrometry}
\label{planetdetection}

\begin{figure}[h]
\includegraphics[width=0.8\textwidth]{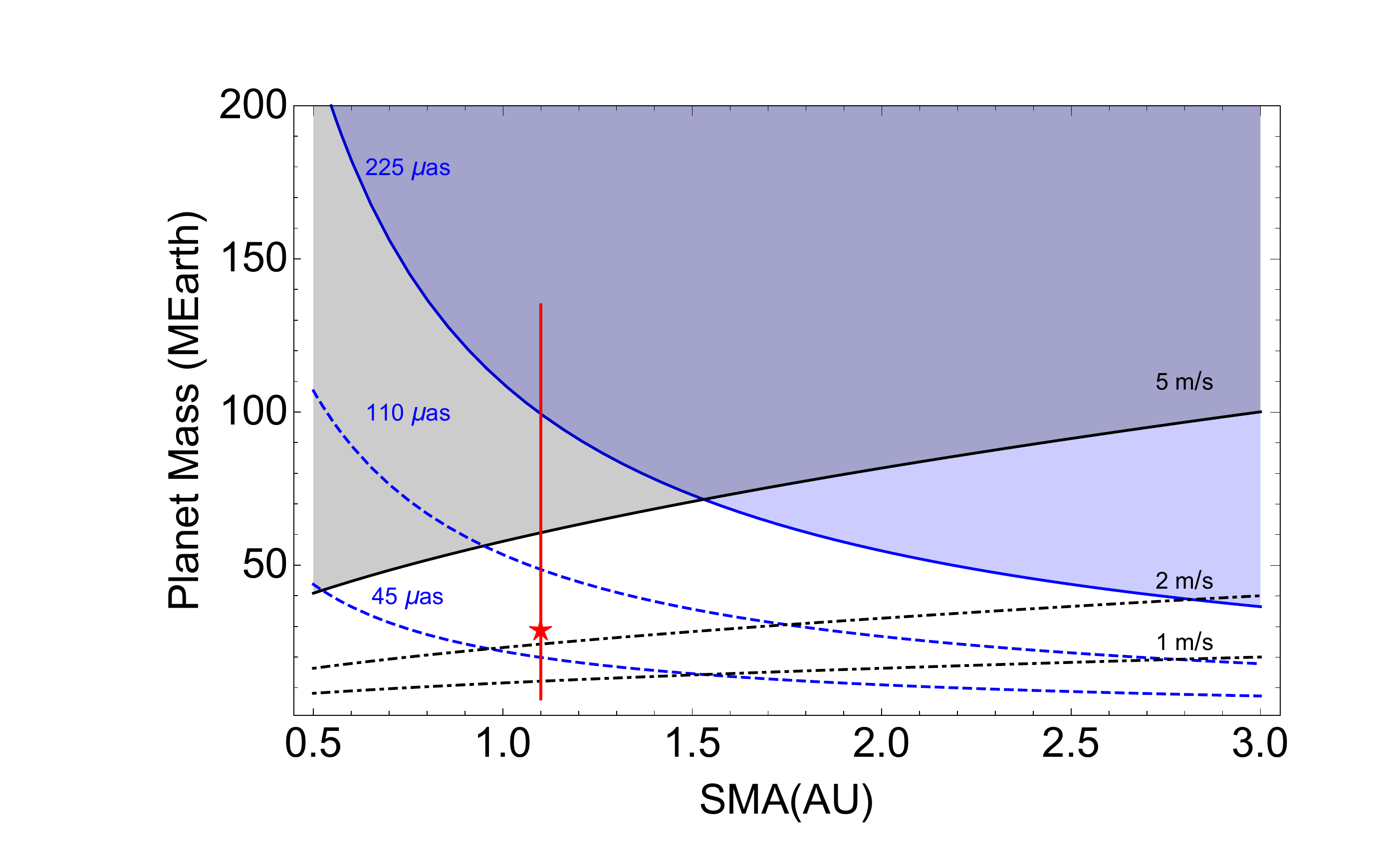}
\caption{The sensitivity of astrometry and radial velocity measurements to planets in  the semi-major axis (SMA) - planet mass plane for three different levels of precision for each technique (PRV in dot-dash black lines and ALMA astrometry in dashed blue lines). In each case we assume 2$\sigma$ limits as described in the text. The red line and star denote the SMA and range of masses corresponding to the candidate planet detected by the VLT NEAR project \citep{Kasper2019,Wagner2020}. \label{compare}}
\end{figure}

Astrometric and PRV measurements have  complementary selection biases. Astrometry favors planets   more distant from the host star, and PRV planets closer-in. Figure~\ref{compare} shows limits to planets in the planet mass vs. semi-major axis plane for the two techniques, each shown with three plausible levels of limiting sensitivity. PRV limits currently have a $\sim2\sigma$ value of 5 m s$^{-1}$ \citep{Zhao2018} which we show potentially improving to 2 and 1 m s$^{-1}$. 
For the ALMA-based astrometry we adopt three cases based on single measurement accuracy of 500, 250, 100 $\mu$as and a series of 20 observations spread over 5 years. The ALMA data presented here demonstrate a differential astrometric accuracy at the 300 $\mu$as level for our longest baseline configuration, that improves as expected with an increasing baseline length.  The longest baselines available at ALMA are $\sim2.5$ times longer than the baselines used for the present observations, thus if the differential astrometry performance continues to scale with baseline length, single epoch observations at $\sim$100 $\mu$as are already feasible.  
To check our detection estimate, we compare with the astrometric data presented by  \citet{Benedict2017}.  They used 31 HST/FGS measurements, spanning 2.9 years, having per observation precision of 800  $\mu$as  to derive a semi-major axis of $760 \pm 110$ $\mu$as (2.4 AU at 46 pc) for HD 202206 c, a brown dwarf with a period of 3.45 yr, resulting in a $\sim 7\sigma$ detection.  With a single measurement precision  of $<$300 $\mu$as (Table~\ref{tab:diffastrom}), an ALMA campaign with similar duration and cadence would yield a noise floor $<50 \,\mu$as. A 20 observation campaign would have a slightly higher noise floor  $\sim 55$ $\mu$as, or a two sigma value of 110 $\mu$as and could lead to the identification of a 25 M$_\oplus$ planet with a 2 AU orbital semi-major axis (Figure~\ref{compare}). 
 
 While such a program would be observationally challenging due to the
 ALMA configuration cycle and a higher SNR would be needed for the
 detection of a planet in the absence of PRV data, ALMA astrometric
 data are clearly relevant in the exoplanet phase space of \acen.  When combined with PRV data, these complementary dynamical measurements will probe the entire 3 AU region around each of the \acen A and B components where planetary orbits are expected to be stable \citep{Quarles2016}.

The mid-infrared direct imaging project NEAR \citep{Kasper2019,Wagner2020} has tentatively  identified a candidate  planet associated with  \acen\ A. While instrument artifacts cannot yet be ruled out, they find an object at 1.1 AU from \acen\ A with a range of possible  radii between 3.3 and 7 M$_\oplus$. Converting radius to mass is of course an uncertain process. Taking transiting planets in this radius range with  masses measured with better than 20\% accuracy\footnote{\url{https://exoplanetarchive.ipac.caltech.edu/index.html} \citep{Akeson2013}} we find an average mass of 30  M$_\oplus$, a minimum mass of 6  M$_\oplus$ and a maximum mass of 135  M$_\oplus$.
While existing measurements would seem to rule out the largest possible masses, future observations of either type could determine  the mass of this putative object.

\subsection{The orbit of Proxima Centauri and \acen\ AB stellar conjunctions}

Our accurate determination of the barycentric parallax, radial velocity, and proper motion of \acen\ AB (Tables~\ref{tab:orbital} and \ref{tab:propermotion}) enables us to improve our knowledge of the relative velocity between the inner AB system and the third component of the system, Proxima Centauri \citep{Kervella2017b}. We adopt the Hipparcos-Gaia proper motion determined by \citet{Kervella2019} for Proxima ($\mu_\alpha = -3781.629 \pm 0.048$\,mas\,a$^{-1}$; $\mu_\delta = +769.421 \pm 0.052$\,mas\,a$^{-1}$) and the radial velocity from \citet{Kervella2017b} ($v_\mathrm{r,abs} = -22.204 \pm 0.032$\,km\,s$^{-1}$. The improved parallax and masses of the AB system translate into an unbound differential velocity limit of $v_\mathrm{max} = 554 \pm 7$\,m\,s$^{-1}$ for a gravitationally bound system. 
We obtain a differential space velocity of $\Delta v_{\alpha-\mathrm{Prox}} = 280 \pm 32$\,m\,s$^{-1}$ between \acen\ AB and Proxima, very significantly below the unbound velocity. The significance of the gravitational link between AB and Proxima has therefore increased over the last three years from $4.4\sigma$ in \citep{Kervella2017b}, to $5.5\sigma$ in \citet{Kervella2019} to $8.3\sigma$ in the present work ($<10^{-15}$ false alarm probability). Proxima becomes a yet more valuable check on lower main sequence stellar modeling. We list the refined parameters for the orbit of Proxima  in Table~\ref{tab:proximaorbit}.

\begin{deluxetable}{lcl}
\tabletypesize{\small}
\tablewidth{0pt}
\tablecaption{Orbital parameters of Proxima.  \label{tab:proximaorbit}}
\tablehead{\colhead{Parameter} & \colhead{Value} &  \colhead{Unit}}
\startdata
Semi-major axis $a$ & $8.2^{+0.4}_{-0.3}$ & kAU \\ 
 Excentricity    $e$ & $0.497^{+0.057}_{-0.060}$ & \\ 
 Period          $P$ & $511^{+41}_{-30}$ & kyr \\ 
 Inclination     $i$ & $124.9^{+2.9}_{-3.2}$ & $\deg$ \\ 
 Longitude of asc. node $\Omega$   & $165^{+3}_{-3}$ & $\deg$ \\ 
 Argument of periastron $\omega$   & $151.0^{+5.7}_{-4.9}$ & $\deg$ \\ 
 Epoch of periastron $T_0$$^a$ & $+278^{+36}_{-28}$ & kyr \\ 
 Periastron radius & $4.1^{+0.7}_{-0.6}$ & kAU \\ 
 Apastron radius   & $12.3^{+0.2}_{-0.1}$ & kAU \\ 
 \enddata
\tablenotetext{a}{The epoch of periastron passage $T_0$ is relative to present.}
\end{deluxetable}

\begin{figure}[h]
\centering
\includegraphics[width=0.7\textwidth]{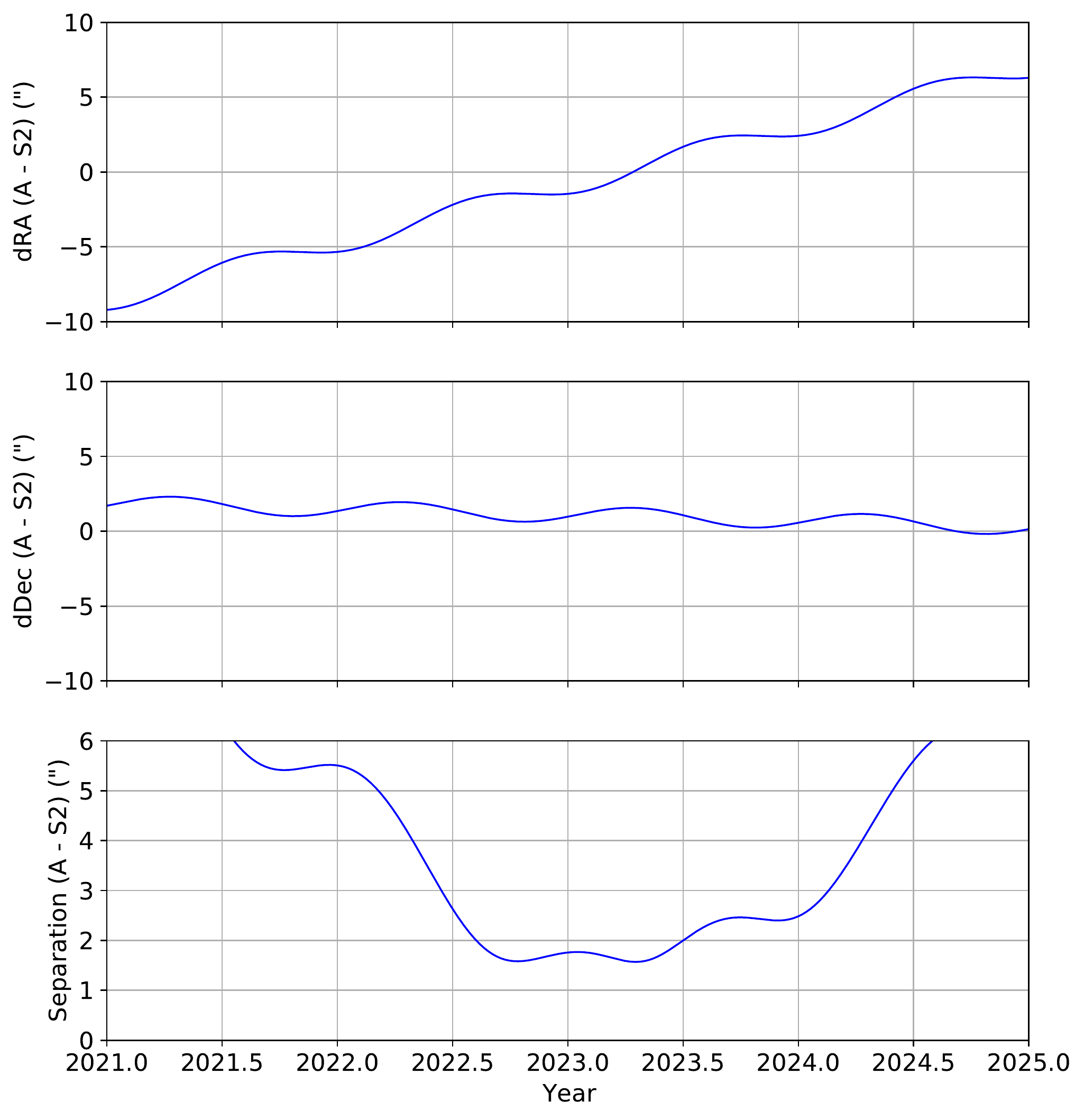}
\caption{The trajectory of the star denoted S2 \citep{Kervella2016} as it approaches \acen\ A. The separation in RA, Dec and total separation is shown as a function of time and incorporates the effects of proper motion, parallax and orbital motion of \acen\ A from this paper and the parallax and proper motion of S2 from Gaia. \label{S2}}
\end{figure}

Using the improved proper motion and orbital parameters of \acen\ AB determined in the present work, it is possible to refine the predictions of the stellar conjunctions between \acen\ AB and background field stars that were identified by K16.
We took into account the positions and proper motions of stars S2, S4 and S6 (see Table 3; K16) that are listed in the Gaia DR3 catalog \citep{gaiadr3}. For stars S1, S3 and S5, we adopted the uncertainties from K16, in particular an uncertainty on their proper motion of $\pm 10$\,mas\,yr$^{-1}$. Thanks to the new ALMA astrometry, the trajectory of \acen\ AB is significantly better constrained, resulting in a significant reduction of the uncertainties on the conjunction parameters in particular the minimum approach angular separation $\rho_\mathrm{min}$.  The trajectory of S2 is shown in Figure~\ref{S2} with a closest approach to \acen\ A of 1\farcs5  in mid 2023. With  $\Delta$K=12.6 mag, S2 will provide a well defined  ``test particle" for searches for Jovian-sized  planets via infrared direct imaging experiments \citep{Beichman2020}.

For stars S1, S3 and S5, the error budget  is dominated by the uncertainty on the position and proper motion of these field stars, and the error bars on $\rho_\mathrm{min}$ are only moderately improved compared to K16 (their Table 3). The revised $\rho_\mathrm{min}$ and dates of closest approach are generally close to the previously determined values, except for the S3 event in 2023 that occurs about six months earlier than calculated in K16.

\begin{deluxetable}{cccccc}
\tabletypesize{\small}
\tablewidth{0pt}
\tablecaption{Conjunctions of \acen\ AB until 2031 with background stars S1 to S6 identified by \citet{Kervella2016} (their Table 3). The minimum separation $\rho_\mathrm{min}$, closest approach date and $\Delta$V magnitude (from K16) are listed. \label{tab:conjunctions}}
\tablehead{\colhead{\acen} & \colhead{$\rho_\mathrm{min}$} &  \colhead{Star} & \colhead{$\Delta$V} & \colhead{Date} & \colhead{Date }\\
 & (arcsec) & & mag & & (decimal year)}
\startdata
B & $0.088 \pm 0.131$ & S01 & 18.7 & 2021-04-20 & 2021.301 \\
 A & $1.509 \pm 0.002$ & S02 & 15.7 & 2023-04-13 & 2023.279\\
 B & $1.484 \pm 0.074$ & S03 & 18.4 & 2023-12-12 & 2023.946\\
 B & $2.283 \pm 0.004$ & S04 & 16.2 & 2024-10-26 & 2024.818\\
 A & $0.170 \pm 0.131$ & S05 & 21.5 & 2028-04-20 & 2028.301\\
 B & $0.294 \pm 0.006$ & S06 & 15.6 & 2031-05-23 & 2031.389\\
  \enddata
\end{deluxetable}

\section{Conclusions}
\label{sec:conclusions}

\begin{enumerate}
\item We have demonstrated that ALMA produces for the nearby bright stars \acen\ A and B absolute astrometric measures with precision and accuracy of order 3 mas and differential separation uncertainties of 300 to 600 $\mu$arcsec.

\item By comparison with the Sun we establish an astrometric jitter due to stellar activity of less than 15 $\mu$arcsec.

\item The combination of historical measurements of \acen\ A-B position angle and separation, more recent PRV measurements from HARPS, and absolute astrometry from {\it Hipparcos},  and ALMA yields improved \acen\ A and B orbital elements, system proper motion and parallax, and component masses.

\item We find stellar parameters of $m_{\rm A} =1.0788 \pm 0.0029$ $M_\odot$ and $m_{\rm B} = 0.9092 \pm 0.0025$ $M_\odot$ and $R_{\rm A} =1.2175 \pm 0.0055$ $R_\odot$ and $R_{\rm B} = 0.8591 \pm 0.0036$ $R_\odot$.  These masses and radii, now improved with a more precise and accurate parallax, will serve to further constrain stellar evolutionary models and provide 'ground truth' for asteroseismological predictions.

\item Our accurate determination of the \acen\ AB system parallax, RV, and proper motion, when compared to those values for Proxima Centauri, confirms that the three stars constitute a bound system.

\item Continued astrometric monitoring of the \acen\ AB system with ALMA, particularly with longer baseline configurations and in combination with precision radial velocity measurements, should reduce planetary companion detection limits to the 
few 10's of M$_\oplus$ range across the entire $<$3 AU region expected to be stable in the \acen\ AB system. 
 \end{enumerate}

\acknowledgments

This paper makes use of the following ALMA data: ADS/JAO.ALMA\#2018.1.00557.S. ALMA is a partnership of ESO (representing its member states), NSF (USA) and NINS (Japan), together with NRC (Canada), MOST and ASIAA (Taiwan), and KASI (Republic of Korea), in
cooperation with the Republic of Chile. The Joint ALMA Observatory is operated by ESO, AUI/NRAO and NAOJ.  The National Radio Astronomy Observatory is a facility of the National Science Foundation operated under cooperative agreement by Associated Universities, Inc. Some of the research described in this publication was carried out in part at the Jet Propulsion Laboratory, California Institute of Technology, under a contract with the National Aeronautics and Space Administration. 
GFB thanks McDonald Observatory for extending a Research Fellow position, allowing continued possession of an office and computing equipment. We thank the referees for a careful reading of the paper and for making many useful suggestions to clarify the material.

\vspace{5mm}
\facilities{ALMA, ESO:3.6m(HARPS)}

\software{CASA (McMullin et al. 2007),
astropy (The Astropy Collaboration 2013, 2018)}

\appendix

\section{Ephemeris of $\alpha$ Cen A and B}

\begin{figure}[h]
\includegraphics[width=0.8\textwidth]{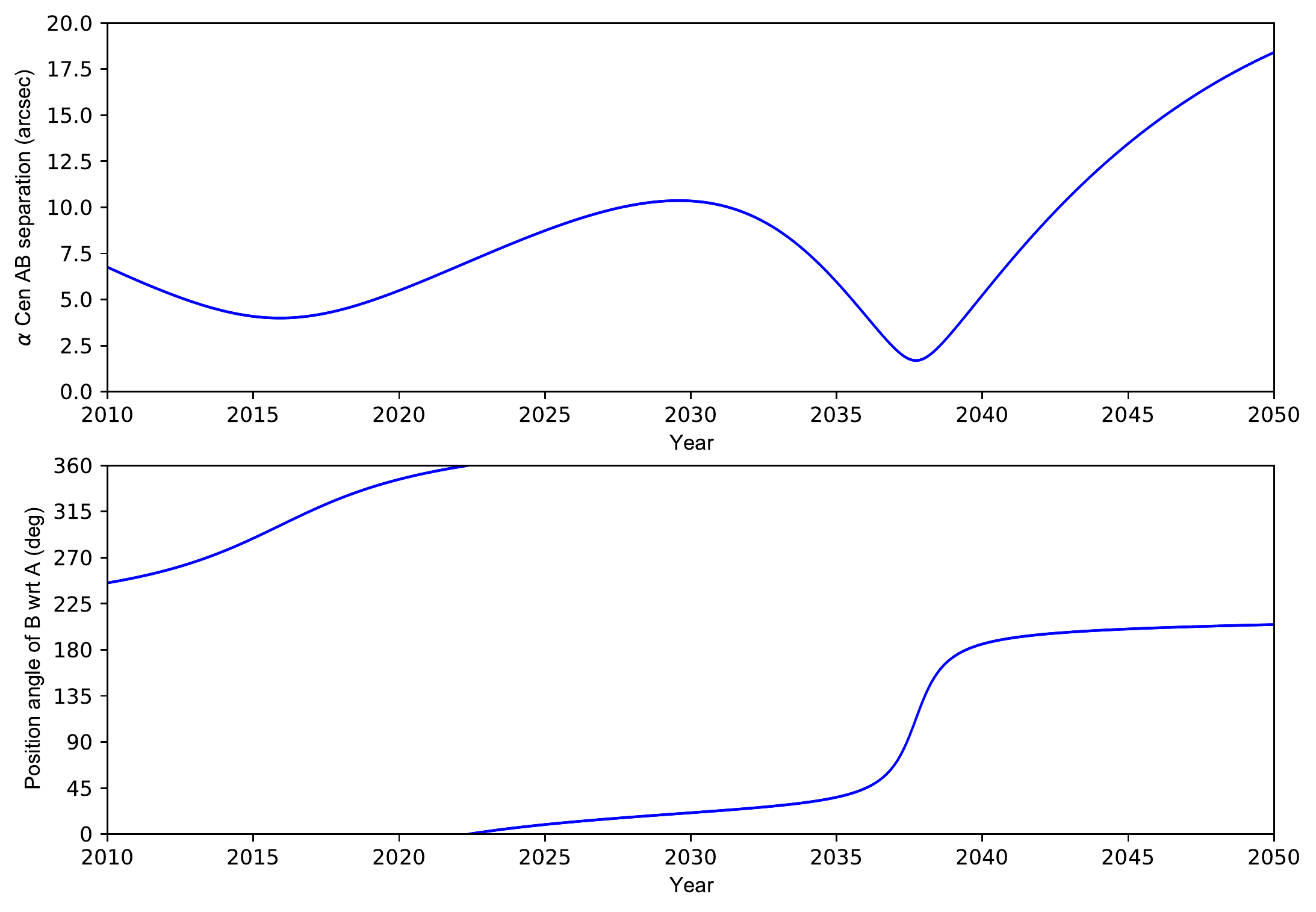}
\caption{Apparent separation and position angle ($N=0, E=90\deg$) of \acen\ B relative to A between 2010 and 2050. \label{seppa}}
\end{figure}

To help in the preparation of future observations of \acen, we present in this section the coordinates of \acen\ A and B as a function of time for the period between 2010 and 2050.
Table~\ref{tab:ephemAB} gives the positions of the AB center of mass, A and B in the ICRS frame. These coordinates take into account the proper motion, orbital motion and parallactic wobble of the two stars. The relative position of B with respect to A in Cartesian and spherical coordinates is also provided, and the evolution of the separation and position angle of \acen\ B relative to A are shown in Fig.~\ref{seppa}.

\begin{deluxetable}{lllllllllll}
\tabletypesize{\tiny}
\tablewidth{0pt}
\tablecaption{Ephemeris of \acen\ A and B between 2010 and 2050. The RA$_{\rm AB}$ and Dec$_{\rm AB}$ columns correspond to the center of mass position.  dRA and dDec are the position offsets of B relative to A, expressed in arcseconds, $\rho$ is the $A-B$ separation, and $\theta$ the position angle of B  with respect to A $(N=0, E=90\deg)$. These coordinates include the proper motion, orbital motion and parallactic wobble of the two stars, and the light time propagation delay. The full table is available in machine-readable format. \label{tab:ephemAB}}
\tablehead{
\colhead{Date} & \colhead{RA$_{\rm AB}$} & \colhead{Dec$_{\rm AB}$} & \colhead{RA$_{\rm A}$} & \colhead{Dec$_{\rm A}$} & \colhead{RA$_{\rm B}$} & \colhead{Dec$_{\rm B}$} & \colhead{dRA} & \colhead{dDec} &  \colhead{$\rho$} & \colhead{$\theta$} \\
&&&&&&& \colhead{arcsec} & \colhead{arcsec} & \colhead{arcsec} & \colhead{degrees}
}
\startdata
2010.00 & 14:39:30.9533 & -60:50:00.386 & 14:39:31.3371 & -60:49:59.098 & 14:39:30.4978 & -60:50:01.913 &  -6.136 &  -2.815 &  6.751 &  245.357 \\ 
2010.05 & 14:39:30.9404 & -60:50:00.527 & 14:39:31.3230 & -60:49:59.259 & 14:39:30.4865 & -60:50:02.032 &  -6.115 &  -2.774 &  6.715 &  245.600 \\ 
2010.10 & 14:39:30.9180 & -60:50:00.665 & 14:39:31.2992 & -60:49:59.415 & 14:39:30.4656 & -60:50:02.148 &  -6.094 &  -2.733 &  6.679 &  245.845 \\ 
2010.15 & 14:39:30.8857 & -60:50:00.782 & 14:39:31.2657 & -60:49:59.551 & 14:39:30.4349 & -60:50:02.243 &  -6.073 &  -2.692 &  6.643 &  246.093 \\ 
2010.20 & 14:39:30.8445 & -60:50:00.863 & 14:39:31.2231 & -60:49:59.651 & 14:39:30.3953 & -60:50:02.302 &  -6.052 &  -2.651 &  6.607 &  246.344 \\ 
2010.25 & 14:39:30.7961 & -60:50:00.897 & 14:39:31.1734 & -60:49:59.704 & 14:39:30.3484 & -60:50:02.313 &  -6.031 &  -2.610 &  6.571 &  246.598 \\ 
2010.30 & 14:39:30.7429 & -60:50:00.878 & 14:39:31.1189 & -60:49:59.703 & 14:39:30.2968 & -60:50:02.272 &  -6.010 &  -2.569 &  6.536 &  246.855 \\ 
2010.35 & 14:39:30.6877 & -60:50:00.805 & 14:39:31.0623 & -60:49:59.649 & 14:39:30.2431 & -60:50:02.176 &  -5.988 &  -2.528 &  6.500 &  247.114 \\ 
2010.40 & 14:39:30.6334 & -60:50:00.682 & 14:39:31.0067 & -60:49:59.545 & 14:39:30.1904 & -60:50:02.031 &  -5.967 &  -2.487 &  6.465 &  247.376 \\ 
2010.45 & 14:39:30.5828 & -60:50:00.519 & 14:39:30.9548 & -60:49:59.400 & 14:39:30.1414 & -60:50:01.846 &  -5.946 &  -2.446 &  6.429 &  247.641 \\ 
2010.50 & 14:39:30.5382 & -60:50:00.327 & 14:39:30.9089 & -60:49:59.227 & 14:39:30.0984 & -60:50:01.632 &  -5.925 &  -2.405 &  6.394 &  247.909 \\ 
2010.55 & 14:39:30.5015 & -60:50:00.121 & 14:39:30.8709 & -60:49:59.040 & 14:39:30.0633 & -60:50:01.404 &  -5.903 &  -2.364 &  6.359 &  248.180 \\ 
2010.60 & 14:39:30.4739 & -60:49:59.918 & 14:39:30.8418 & -60:49:58.856 & 14:39:30.0372 & -60:50:01.178 &  -5.882 &  -2.322 &  6.324 &  248.454 \\ 
2010.65 & 14:39:30.4555 & -60:49:59.733 & 14:39:30.8221 & -60:49:58.689 & 14:39:30.0204 & -60:50:00.971 &  -5.861 &  -2.281 &  6.289 &  248.730 \\ 
2010.70 & 14:39:30.4459 & -60:49:59.580 & 14:39:30.8112 & -60:49:58.555 & 14:39:30.0124 & -60:50:00.795 &  -5.839 &  -2.240 &  6.254 &  249.010 \\ 
2010.75 & 14:39:30.4437 & -60:49:59.471 & 14:39:30.8076 & -60:49:58.466 & 14:39:30.0118 & -60:50:00.665 &  -5.818 &  -2.199 &  6.219 &  249.294 \\ 
2010.80 & 14:39:30.4468 & -60:49:59.415 & 14:39:30.8094 & -60:49:58.428 & 14:39:30.0165 & -60:50:00.586 &  -5.796 &  -2.158 &  6.185 &  249.580 \\ 
2010.85 & 14:39:30.4525 & -60:49:59.413 & 14:39:30.8138 & -60:49:58.445 & 14:39:30.0239 & -60:50:00.562 &  -5.775 &  -2.117 &  6.150 &  249.869 \\ 
2010.90 & 14:39:30.4579 & -60:49:59.464 & 14:39:30.8178 & -60:49:58.515 & 14:39:30.0308 & -60:50:00.591 &  -5.753 &  -2.075 &  6.116 &  250.162 \\ 
2010.95 & 14:39:30.4597 & -60:49:59.560 & 14:39:30.8182 & -60:49:58.629 & 14:39:30.0342 & -60:50:00.663 &  -5.731 &  -2.034 &  6.082 &  250.458 \\ 
2011.00 & 14:39:30.4552 & -60:49:59.686 & 14:39:30.8124 & -60:49:58.775 & 14:39:30.0314 & -60:50:00.768 &  -5.710 &  -1.993 &  6.048 &  250.758 \\ 
2011.05 & 14:39:30.4424 & -60:49:59.827 & 14:39:30.7982 & -60:49:58.935 & 14:39:30.0201 & -60:50:00.887 &  -5.688 &  -1.952 &  6.014 &  251.061 \\ 
2011.10 & 14:39:30.4199 & -60:49:59.965 & 14:39:30.7744 & -60:49:59.092 & 14:39:29.9993 & -60:50:01.002 &  -5.666 &  -1.911 &  5.980 &  251.367 \\ 
... \\
\enddata
\end{deluxetable}

\bibliographystyle{aasjournal}

\end{document}